\documentclass[showpacs,aps,pra,reprint,amsmath,amssymb]{revtex4-1}
\usepackage{amsmath}
\usepackage{epsfig}
\usepackage{subfigure}
\usepackage{graphicx}
\usepackage{dcolumn}
\usepackage{stmaryrd}
\usepackage{mathrsfs}
\usepackage{pifont}
\usepackage{amsthm}
\usepackage{amssymb}
\usepackage{bm}
\usepackage{latexsym}
\usepackage{hyperref}
\usepackage{color}
\usepackage{amsfonts}%
\usepackage{textcomp}
\providecommand{\U}[1]{\protect\rule{.1in}{.1in}}
\newcommand{\la}{\langle}
\newcommand{\ra}{\rangle}
\newcommand{\beq}{\begin{equation}}
\newcommand{\eeq}{\end{equation}}
\newcommand{\beqa}{\begin{eqnarray}}
\newcommand{\eeqa}{\end{eqnarray}}
\newcommand{\lam}{\lambda}
\newcommand{\+}{^\dagger}

\newcommand{\si}{\sigma}

\newcommand{\Om}{\Omega}
\newcommand{\om}{\omega}
\newcommand{\de}{\delta}
\newcommand{\be}{\beta}

\newcommand{\pa}{\partial}
\newcommand{\app}{\approx}
\newcommand{\non}{\nonumber}
\newcommand{\dg}{\mathcal{D}_g[\bx]}

\newcommand{\ri}{{\rm I}}
\newcommand{\rs}{{\rm S}}
\newcommand{\rr}{{\rm E}}

\newcommand{\rt}{{\rm tot}}
\newcommand{\eq}{\equiv}
\newcommand{\sq}{\sqrt}
\newcommand{\ora}{\overrightarrow}

\newcommand{\raw}{\rightarrow}

\newcommand{\bx}{{\bm \xi}}
\newcommand{\mm}{{\mathcal M}}
\newcommand{\hh}{\hat{H}}
\newcommand{\hb}{\hat{b}}
\newcommand{\hc}{\hat{c}}
\newcommand{\hd}{\hat{d}}
\newcommand{\hq}{\hat{Q}}
\newcommand{\hp}{\hat{P}}
\newcommand{\hrr}{\hat{\rho}_r}

\newcommand{\hv}{\hat{v}}
\newcommand{\bq}{\bar{Q}}
\newcommand{\ua}{\uparrow}
\newcommand{\da}{\downarrow}

\newcommand{\rhc}{{\rm H.c.}}
\newcommand{\Tr}{{\rm Tr}}
\newcommand{\rv}{{\rm vac}}
\newcommand{\fr}{\frac}

\newcommand{\q}{\quad}
\newcommand{\pr}{\prod}
\newcommand{\cs}{\cdots}
\newcommand{\hN}{\hat{N}}

\newcommand{\Lam}{\Lambda}
\newcommand{\vDe}{\varDelta}
\newcommand{\ift}{\infty}
\newcommand{\6}{\!\!\!\!\!\!}
\newcommand{\3}{\!\!\!}
\newcommand{\ze}{^{\{0\}}}
\newcommand{\e}{\!\!}
\newcommand{\ep}{\epsilon}
\newcommand{\ta}{\tau}

\begin{document}

\title{Open system approach to non-equilibrium dynamical theory of quantum dot systems}

\author{Wufu Shi$^{1,2}$}
\email{wshi1@stevens.edu}

\author{Yusui Chen$^{3,1}$}
\email{ychen132@nyit.edu}

\author{Lihui Sun$^{1,4}$}
\email{lhsun@yangtzeu.edu.cn}

\author{J. Q. You$^{5}$}

\author{Ting Yu$^{1,2}$}
\email{Corresponding Author: tyu1@stevens.edu}

\affiliation{$^1$Center for Quantum Science and Engineering and Department of Physics, Stevens Institute of
Technology, Hoboken, New Jersey 07030, USA}

\affiliation{$^{2}$ Beijing Computational Science Research Center, Beijing 100084, China}

\affiliation{$^3$Department of Physics, New York Institute of Technology, Old Westbury, NY11568, USA}

\affiliation{$^4$Institute of Quantum Optics and Information Photonics,
	School of Physics and Optoelectronic Engineering, Yangtze University, Jingzhou
	434023, China}

\affiliation{$^5$Zhejiang Province Key Laboratory of Quantum Technology and Device, Zhejiang University, Hangzhou 310027, China}

\date{\today }

\begin{abstract}
We theoretically investigate the non-equilibrium quantum dynamical theory of a quantum dot system coupled to fermionic reservoirs
using the recently developed stochastic fermionic quantum state diffusion (FQSD) equation. The exact or approximate 
dynamical equations associated with the FQSD equation can describe the non-equilibrium quantum transport processes 
beyond the long-time limit leading to a steady state. We study in details the electron transport of a quantum-dot system 
coupled to two fermionic environments with different chemical
potentials. We report the onset of Coulomb blockade in quantum dots in two distinctive cases: one involving a spin degeneracy one-quantum dot model, 
and the other a specific spin non-degeneracy two-quantum dot model.  While the spin degeneracy case shows that the current in the quantum dot 
may be blockaded non-monotonically with respect to Coulomb energy, the non-degeneracy case exhibits significant non-Markovian effects, and it enables us to study the relations between initial conditions of the dots and the steady state currents.

%
\end{abstract}

\pacs{03.65.Yz, 05.30.Fk, 42.50.Lc, 05.40.-a}
\maketitle

\section{Introduction}

As a potential candidate of quantum devices in the implementation of quantum information processing~\cite{Loss,Nori1,Nori2,Ladd,Nori3},
quantum dot systems and their quantum transport phenomena have been studied extensively from different points of view~\cite{Meir,Wiseman,Ventra}. 
From a viewpoint of quantum open system, the source and drain in the electron transport may be modelled by two thermal equilibrium reservoirs with two different chemical potentials. Traditionally, treatments of the interaction of a small quantum system with an environment are based on a variety of 
approaches, such as the many-body Schr\"{o}dinger 
equation, the non-equilibrium Green's function method~\cite{xx, Hu0},  the input-output approach~\cite{InputOutput}, the path integral approach~\cite{hu,Feynman-Vernon, Leggett, Hu1}, and the stochastic Schr\"{o}dinger equation (SSE) method~\cite{Gisin,QuanJump1,QuanJump2,QuanJump3,Diosi1,Diosi2,Strunz,Jing}. 
More specifically, the SSE method represented by a non-Markovian quantum state diffusion (QSD) equation can simulate the non-Markovian features of the open quantum systems arising in many interesting scenarios such as a structured environment, a time-delayed
external control, or the strong system-environment coupling etc. 
Such a non-Markovian approach will be useful for studying the temporal behaviour of quantum transport processes in different time scales that is of interest from recent attempts in understanding  quantum decoherence and non-equilibrium dynamics. Several approaches have been studies recently
including a non-Markovian master equation (ME) for quantum dots coupled to fermionic non-Markovian environments derived by
using the Feynman-Vernon influence functional (IF) approach~\cite{ZhangDQD}.  A different treatment based on the fermionic non-Markovian stochastic Schr\"{o}dinger equation (NMSSE) method has been 
developed~\cite{Shi2013c, Zhao2012d, Chen2013}, and it has been demonstrated that the exact ME can 
be derived for a quadratic Hamiltonian. For a system with a more generic non-quadratic Hamiltonian, a systematic perturbation is available for numerical calculations. 

The purpose of this paper is  to study quantum dynamical processes of a class of quantum dot systems coupled to one or more fermionic reservoirs 
by using  the perturbative NMSSE method. We use the NMSSE method~\cite{Shi2013c, Zhao2012d, Chen2013} to derive 
non-Markovian MEs for the density operators of quantum dot systems coupled to their fermionic environments.
We show, in particular, that the NMSSE with quartic interaction terms (Coulomb interaction terms) plays an important role in describing a non-equilibrium process such as the temporal behaviours of the quantum transport processes.  
We show that the transport processes with the Coulomb interaction can be studied with an approximate ME, and in case of the absence of the Coulomb interaction, an exact ME can be obtained from NMSSE in a straightforward manner.
It should be noted that the Coulomb blockade effects are studied for both spin degeneracy and non-degeneracy cases.  
In the latter case, we report the observation of a correlated noise arising from two separate fermionic reservoirs.

The paper is organized as follows. In Sec.~\ref{Transport}, we begin with a brief introduction of the NMSSE and a generic non-Markovian ME 
derived from the NMSSE for the density operator of a single quantum dot coupled to the source and drain modelled by two fermionic reservoirs. 
Then we discuss the effect of Coulomb blockade in two model systems in Sec.~\ref{Coulomb}. In the spin degeneracy one-quantum dot model, 
we investigate the influence of the bandwidth and chemical potential on the Coulomb blockade effect. In addition,
in the spin non-degeneracy two-dot model, with the two-quantum dot resonance condition we show that the steady state current 
is sensibly dependent on the initial states of the quantum dots. Our discussions and final comments are presented in Sec.~\ref{conclusion}.  
Appendix A gives the details  of the numerical algorithm for the evaluation of the coefficients of the exact ME in Sec.~\ref{Transport}.

\section{Transport Between Two environments through a quantum dot}
\label{Transport}

\begin{figure}[ht]
\begin{center}
\includegraphics[width=2.6in]{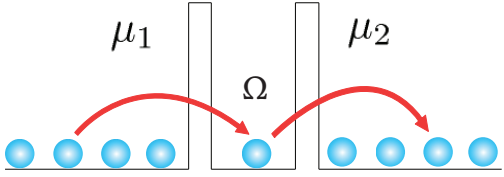}
\end{center}
\caption{A schematic diagram of a single quantum dot with electrons
transported between two environments.}
\label{figcoulomb2}
\end{figure}

To begin with, we consider a simple system composed of a single quantum dot coupled to two fermionic
environments which correspond to a source and a drain, respectively, as shown in Fig.~\ref{figcoulomb2}. 
The environments may represent electrodes or impurities.
We assume that the potential barriers between the quantum dot and environments
can localize the electron's mode in the dot, but still permits the electron tunnel processes. 
In this case, the Hamiltonian of the total system can be written in the framework of
system-plus-environment (we set $\hbar =1$ throughout the paper):
 \begin{eqnarray}
 \hh_\rt = \hh_\rs +\hh_\ri +\hh_\rr, \label{t1}
 \end{eqnarray}
 where $\hh_\rs =  \Om\hd\+\hd$ is the Hamiltonian of the quantum dot,
 $\hh_\rr = \sum_k \om_k (\hb'_{1k}{}\+\hb'_{1k} +\hb'_{2k}{}\+\hb'_{2k})$ is the Hamiltonian of the source and drain, and
 $\hh_\ri =  \sum_k \sum_{j=1,2}t_{jk} \left( \hd\+\hb_{jk}' + \hb_{jk}'{}\+\hd\right)$
 is the interaction Hamiltonian between the dot and the environments (source and drain). Here, $\hd\+$ and $\hd$ are respectively the fermionic creation 
 and annihilation operators of the electrons in the quantum dot, and $\hb'{}\+_{jk}$ and $\hb'_{jk} $ are respectively the fermionic
creation and annihilation operators of the $k$th mode of the $j$th environment ($j=1,2$), and $t_{jk}$ stands for
the coupling strength between the quantum dot and the $k$th mode of the $j$th environment. The fermionic operators satisfy 
the well-known anticommutation relations: $\{\hd\+,\hd\}=1$, $\{\hb'{}\+_{jk},\hb'_{j'k'} \} =\de_{kk'}\de_{jj'}$, and 
$\{\hb'{}\+_{jk},\hd \} =\{\hd\+,\hb'_{jk} \} =0$. For simplicity, we have assumed that $t_{jk}$ are real numbers without losing generality.

The initial states of the environments are set as two thermal equilibrium states at the same temperature $T$, but with different
chemical potentials $\mu_j$. 
The NMSSE method for the thermal environments can be treated by using a Bogoliubov transformation to convert a thermal environment problem to 
that for a vacuum environment. For this purpose, we first decompose the thermal state into an entangled pure state: 
\begin {eqnarray}
 && (1-\bar{n}_{jk})|0_{b_{jk}'}\ra\la0_{b_{jk}'}| +\bar{n}_{jk}|1_{b_{jk}'}\ra\la1_{b_{jk}'}|  \non \\
 \raw\; && \sq{1-\bar{n}_{jk}}|0_{b'_{jk}},0_{c'_{jk}}\ra +\sq{\bar{n}_{jk}}|1_{b_{jk}'},1_{c_{jk}'}\ra.  \label{purification}
\end {eqnarray}
$\bar{n}_{jk}$ is the average number of electrons in the $k$th mode of the $j$th environment $\om_k\hb'_{jk}{}\+\hb^{\prime}_{jk}$, 
which satisfies the Fermi-Dirac distribution $\bar{n}_{jk}=1/(1+e^{\be(\om_k -\mu_j)})$, here $\be$ is the inverse temperature
$\be=1/(k_BT)$. $c'_{jk}$ is a fictitious mode corresponding to the negative
energy: $-\om_k\hc'_{jk}{}\+\hc'_{jk}$ which may be understood as a electron hole. The introduction of those electron-hole pairs can essentially
transform the finite-temperature problem into a zero-temperature one.  The hole system introduced this way is not interacting with the electron 
reservoirs, so it does not affect the system dynamics, but simply serves as ancilla qubits for purification of the environments. In fact, if we trace out all the fictitious modes of the entangled pure state,
the thermal state of $b'_{jk}$ can be obtained.
Then we apply a  Bogoliubov transformation to convert such an entangled pure state into a vacuum. More specifically,  if we choose the following  transformation:
 \begin{eqnarray}
  \hat{b}_{jk} &=& \sq{1-\bar{n}_{jk}}\,\hat{b}_{jk}'-\sq{\bar{n}_{jk}}\,\hat{c}_{jk}'{}\+,  \non\\
  \hat{c}_{jk} &=& \sq{1-\bar{n}_{jk}}\,\hat{c}_{jk}'+\sq{\bar{n}_{jk}}\,\hat{b}_{jk}'{}\+,  \label{t5}
\end{eqnarray}
while its inverse is
\begin{eqnarray}
 \hat{b}_{jk}' &=& \sq{1-\bar{n}_{jk}}\,\hat{b}_{jk}+\sq{\bar{n}_{jk}}\,\hat{c}_{jk}\+ ,\non\\
  \hat{c}_{jk}' &=& \sq{1-\bar{n}_{jk}}\,\hat{c}_{jk}-\sq{\bar{n}_{jk}}\,\hat{b}_{jk}\+,  \label{t6}
 \end{eqnarray}
then the entangled state in Eq.~(\ref{purification}) may be converted to a vacuum corresponding to the annihilation operators $\hb_{jk}$ and 
$\hc_{jk}$:
\begin {eqnarray}
 \sq{1-\bar{n}_{jk}}|0_{b'_{jk}},0_{c'_{jk}}\ra +\sq{\bar{n}_{jk}}|1_{b_{jk}'},1_{c_{jk}'}\ra \raw |0_{b_{jk}},0_{c_{jk}}\ra. \q
\end {eqnarray}

Following the transformation, the Hamiltonian (\ref{t1}) the takes the following form:
 \begin {eqnarray}
  \hh_\rt' &=& \Om\hd\+\hd + \sum_k \sum_{j=1,2} \om_k \left(\hb_{jk}\+\hb_{jk} - \hc_{jk}\+\hc_{jk} \right) \non \\
  &+& \sum_k \sum_{j=1,2} (t_{b_{jk}} \hd\+\hb_{jk} +t_{c_{jk}}\hd\+\hc_{jk}\+ +\rhc), \label{t7}
 \end {eqnarray}
 where $t_{b_{jk}} \eq \sq{1-\bar{n}_{jk}}t_{jk}$ and $t_{c_{jk}} \eq \sq{\bar{n}_{jk}}t_{jk}$. 
 Therefore, if we choose the initial state of the environment modes $b_{jk}$ and $c_{jk}$ in the
Hamiltonian (\ref{t7}) to be a vacuum state, then by tracing over all $c^{\prime}_{jk}$ modes we will get
 the Hamiltonian (\ref{t1}) involving only the environment modes $b^{\prime}_{jk}$, which are initially in a thermal state.

In what follows, we will develop an NMSSE approach and define the fermionic quantum trajectory for the quantum dot system. 
The NMSSE approach will involve the Grassmann-Bargmann (GB) coherent state representation ~\cite{Glauber1999}, which
may stimulate  an anti-commutating Gaussian noise. The GB coherent state representation is
defined as
\begin{eqnarray}
\|\bx\ra \equiv \prod_{j=1,2;k} (1-\xi_{b_{jk}} \hb_{jk}\+)(1-\xi_{c_{jk}}\hc_{jk}\+)|{\rm vac}\ra_\rr,
\end{eqnarray}
where $\xi_{b_{jk}}$ and $\xi_{c_{jk}}$ are independent Grassmann variables, and $\bx$
is a collective notation.
A fermionic quantum trajectory can be defined as the inner product of the GB coherent states and the wave function of the total system
$|\psi_\rt^I(t)\ra$:
\begin {eqnarray}
   |\psi_t(\bx^*)\ra &\equiv& \la\bx\|\psi_\rt^I(t)\ra \non\\
   &=& \la\bx \|e^{i \sum_{j=1,2;k} \om_k(\hb_{jk}\+\hb_{jk} -\hc_{jk}\+\hc_{jk} )t}
|\psi_\rt(t)\ra .
\end {eqnarray}
The reduced density operator of the quantum dot system can be obtained by taking the mean value of the fermionic quantum trajectories over the Grassmann variables: 
 \begin {eqnarray}
  \hrr \!=\! \int \!\dg |\psi_t(\bx^*\!)\ra\la\psi_t(-\bx)| \!=\!\mm (|\psi_t(\bx^*\!)\ra\la\psi_t(-\bx)|), \q\q  \label{unravel}
 \end {eqnarray}
where the Grassmann-Gaussian measure is defined as
\begin {eqnarray}
 \dg \eq \pr_{\lam=b,c} \pr_{j=1,2; k} (d\xi_{\lam_{jk}}^* d\xi_{\lam_{jk}} e^{-\xi_{\lam_{jk}}^*\xi_{\lam_{jk}}}),
\end {eqnarray}
and $\mm(\cs)$ stands for the fermionic stochastic mean value. 
The time evolution of the trajectories is governed by the NMSSE:
\begin {eqnarray}
   \pa_t|\psi_t(\bx^*)\ra &=& \Big(-i\hh_\rs +\sum_{j=1,2} (-\hd\+\bq_{b_j}-\hd\xi^*_{b_jt}+\hd\+\xi^*_{c_jt}  \non \\
&& +\hd\bq_{c_j} ) \Big)|\psi_t(\bx^*)\ra,  \label{QSD1}
\end {eqnarray}
where the stochastic processes and Q-operators are defined as follows,
\begin{eqnarray}
&& \xi^*_{b_jt} \eq -i\sum_k t_{b_{jk}} e^{i\om_k t} \xi^*_{b_{jk}}, \; K_{b_j}(t,s) = \sum_k t_{b_{jk}}^2 e^{ -i\om_k(t-s)},  \non\\
&& \xi^*_{c_jt} \eq -i\sum_k t_{c_{jk}} e^{-i\om_k t} \xi^*_{c_{jk}}, \; K_{c_j}(t,s) = \sum_k t_{c_{jk}}^2 e^{ i\om_k(t-s)}, \non\\
&& \bq_{\lam_j}(t,\bx^*) \eq \int_0^t ds K_{\lam_j}(t,s) \hq_{\lam_j}(t,s,\bx^*), \q  (\lam=b,c),  \non\\
 &&\hq_{\lam_j}(t,s,\bx^*)|\psi_t(\bx^*)\ra \eq \ora{\de}_{\xi_{\lam_js}^*}|\psi_t(\bx^*)\ra,  
\end{eqnarray}
 in which  $\ora{\de}_{\xi_{\lam_js}^*}$ means the left-functional-derivative with respect to noise $\xi_{\lam_js}^*$, and $K_{\lam_j}(t,s)$ are
 correlation functions defined as $K_{\lam_j}(t,s) \eq \mm(\xi_{\lam_jt}\xi_{\lam_js}^*)$. Eq.~(\ref{QSD1}) is also 
 called the fermionic quantum state diffusion (FQSD) equation.

When we apply an extended Novikov theorem (for the bosonic case, see Ref.~\cite{YDGS99}) 
to calculate the stochastic mean value:
 $\mm \big( \xi^*_{\lam_jt}\hp \big)$, where $\hp \eq |\psi_t(\bx^*)\ra \la\psi_t(-\bx)|$, 
 a formal non-Markovian ME is derived:
 \begin{eqnarray}
  \pa_t \hrr &=& -i[\hh_\rs, \hrr] +\sum_j \Big( \big[\mm(\bq_{b_j}(t,\bx^*) \hp), \hd\+\big]\! \non \\
  && +\!\big[\hd, \mm(\bq_{c_j}(t,\bx^*)\hp) \big]\!+\!\rhc \Big). \label{fME1}
 \end{eqnarray}
 
Using the Heisenberg approach, the mean value terms above can be exactly evaluated:
\begin {eqnarray}
 && \mm\big(\bq_{b_j}(t,\bx^*)\hp -(\bq_{c_j}(t,\bx^*)\hp)\+ \big)  \non \\
 &=& F_{1j}(t)\hd\hrr(t) +F_{2j}(t)\hrr(t)\hd,  \label{Meanterm}
\end {eqnarray}
where the expressions of $F_{ij}$ can be found in Appendix A. By substituting Eq.~(\ref{Meanterm}) into Eq.~(\ref{fME1}), 
we obtain the ME:
\begin {eqnarray}
 \pa_t \hrr \!=\! -i[\hh_\rs,\hrr] \!+\!\sum_{j} \big([F_{1j}\hd\hrr \!+\!F_{2j}\hrr\hd,\hd\+] +\rhc \big). \q\q
\end {eqnarray}

Analogue to the classical continuity equation: $\pa_t \rho +\nabla\cdot \vec{j} =0$, the current of the quantum dot can be connected to
the change rate of the particle number.  We split the formal ME (\ref{fME1}) into a system part and two environment-involved
parts:
\begin {eqnarray}
 \pa_t \hrr = -i[\hh_\rs, \hrr] +(\hv_{1d} -\hv_{d2}),
\end {eqnarray}
where $\hv_{1d}$ and $\hv_{d2}$ are defined as
\begin {eqnarray}
 \hv_{1d} &\eq& \big([\mm(\bq_{b_1}\hp), \hd\+] \!+\![\hd, \mm(\bq_{c_1}\hp)] +\rhc \big),  \non \\
 \hv_{d2} &\eq& \big([\hd\+\!, \mm(\bq_{b_2}\hp)] \!+\![\mm(\bq_{c_2}\hp), \hd]\! +\rhc \big).  \label{1hv}
\end {eqnarray}
Usually $\hh_\rs$ commutes with the system particle number operator $\hN$, so we have
\begin {eqnarray}
 \pa_t \Tr(\hrr(t) \hN) &=& -i\Tr(\hh_\rs\hrr\hN -\hrr\hh_\rs\hN)  \non \\
 && +\Tr(\hv_{1d}\hN -\hv_{d2}\hN)  \non \\
 &=& \Tr(\hv_{1d}\hN -\hv_{d2}\hN).  
\end {eqnarray}
Corresponding to the $\nabla\cdot \vec{j}$ term, we can define $I_{1d} \eq q_e\Tr \big( \hv_{1d} \hat{N} \big)$, and
$I_{d2} \eq q_e\Tr \big( \hv_{d2} \hat{N} \big)$, which stand for the expectation values of currents with respect to directions:
$1 \raw d$ and $d\raw2$ respectively. Here $1$ and $2$ denote the $1$-st and the $2$-nd environment, $d$ denotes the quantum dot, and
$q_e$ is the charge of one electron.

By comparing Eq.~(\ref{Meanterm}) with Eq.~(\ref{1hv}), the currents may be obtained:
\begin {eqnarray}
 \hv_{1d} &=& [F_{11}\hd\hrr +F_{21}\hrr\hd, \hd\+] +\rhc,   \non \\
 \hv_{d2} &=& [\hd\+, F_{12}\hd\hrr +F_{22}\hrr\hd] +\rhc,  \non \\ 
 I_{1d} &=& -2q_e(F_{21}^R \rho_{00} + F_{11}^R \rho_{11}),  \non \\
 I_{d2} &=& 2q_e(F_{22}^R \rho_{00} + F_{12}^R \rho_{11}),  \label{Current1D2E}
\end {eqnarray}
where the elements $\rho_{00}$ and $\rho_{11}$ are defined as $\hrr \eq \rho_{00}|0\ra\la 0| +\rho_{11}|1\ra\la 1|$, and the
superscript $R$ denotes the real part of a complex number.

\begin{figure}[!htb]
 \centering\includegraphics[scale=0.57]{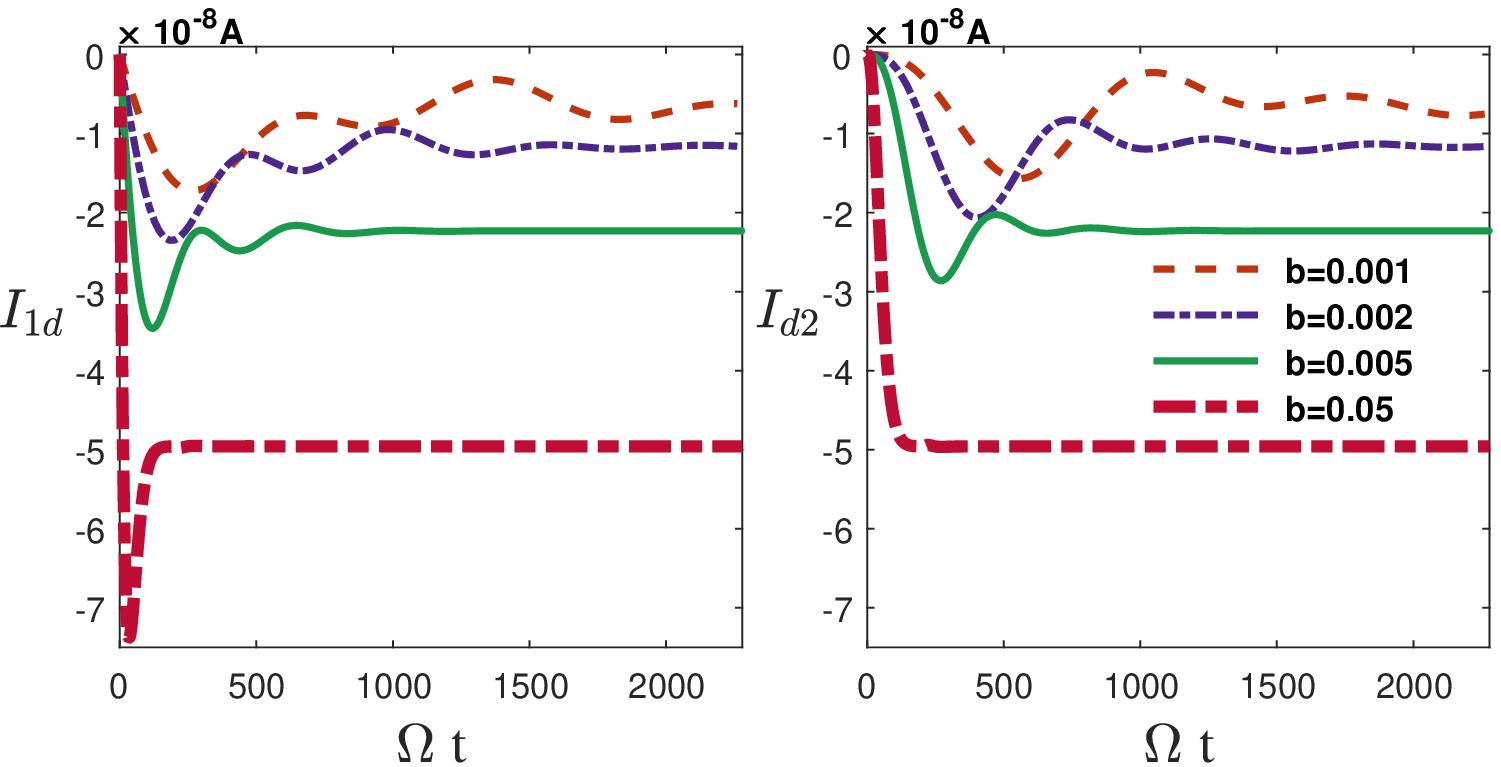}
 \caption{ Single quantum dot interacting with double environments. The initial state of the dot system is
 $\hrr(0) =|\rv_\rs\ra\la\rv_\rs|$. The parameters are set as $\Lam=0.05$, $\vDe\om =2\pi {\rm Grad/s}$,
 $T=2{\rm K}$, $\mu_1=40$${\rm meV}$, $\mu_2=0$ and $\Om=30$${\rm meV}$. } 
 \label{SingleDotDouEnvCurr1}
\end{figure}
\begin{figure}[!h]
 \centering\includegraphics[scale=0.57]{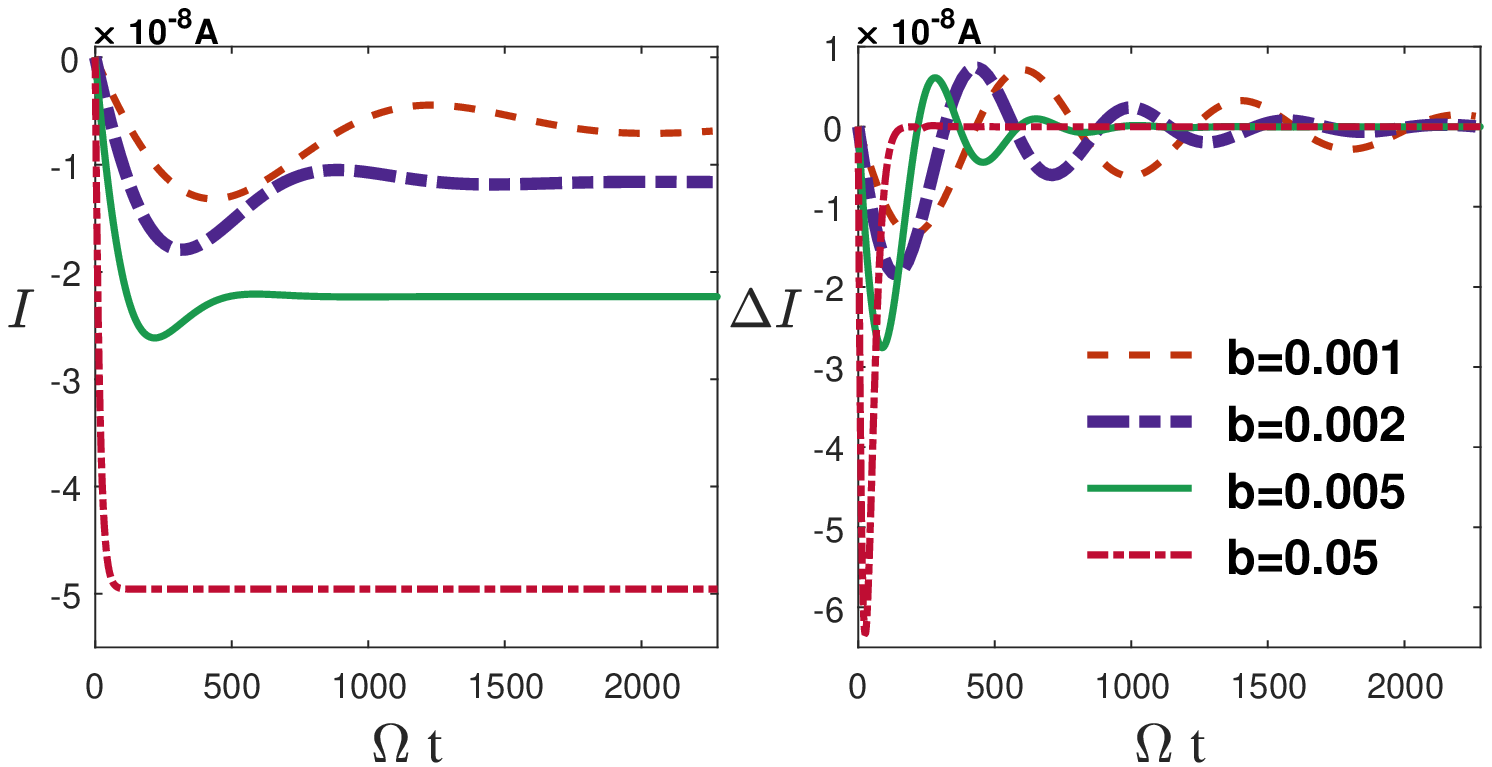}
 \caption{Single quantum dot interacting with double environments. The initial state of the system is:
 $\hrr(0) =|\rv_\rs\ra\la\rv_\rs|$. The parameters are set as $\Lam=0.05$, $\vDe\om =2\pi {\rm Grad/s}$,
 $T=2{\rm K}$, $\mu_1=40$${\rm meV}$, $\mu_2=0$ and $\Om=30$${\rm meV}$. } 
 \label{SingleDotDouEnvCurr2}
\end{figure}

  
For simplicity, we use the Lorentzian spectrum in our numerical simulations:
\begin {eqnarray}
 t_{jk}^2 =\Lam^2\Om\vDe\om \fr{b^2}{(\om_k/\Om -1)^2 +b^2},
\end {eqnarray}
where $\vDe\om$ is the eigen-frequency spacing of the discrete spectrum, 
the two environments are represented by the same relative bandwidth $b$ and the same coupling strength factor $\Lam$. 
In Fig.~\ref{SingleDotDouEnvCurr1}, we can see how the bandwidth $b$ influence the currents in the extreme non-Markovian case.
When $b$ decreases, the amount of the currents reduces to $0$, and the oscillation start to take place. 
The difference between $I_{1d}$ and $I_{d2}$ is related to $\Tr\big(\hd\+\hd\,\pa_t\hrr(t) \big)$, so if $\hrr$ has a steady state,
$\vDe I\eq I_{1d} -I_{d2}$ will converge to $0$. In Fig.~\ref{SingleDotDouEnvCurr2}, we can see that the average current: 
$I\eq(I_{1d}+I_{d2})/2$ is smoother than $I_{1d}$ and $I_{d2}$, 
and display a simpler dissipated oscillation pattern. Actually, in this model $I(t)$ will not be affected by $\hrr(0)$ 
due to the symmetry between the spectrum $t_{1k}$ and $t_{2k}$. 

In Fig.~\ref{SingleDotDouEnvCurr2}, we can see that when the bandwidth
\begin{figure}[!htb]
 \centering\includegraphics[scale=0.57]{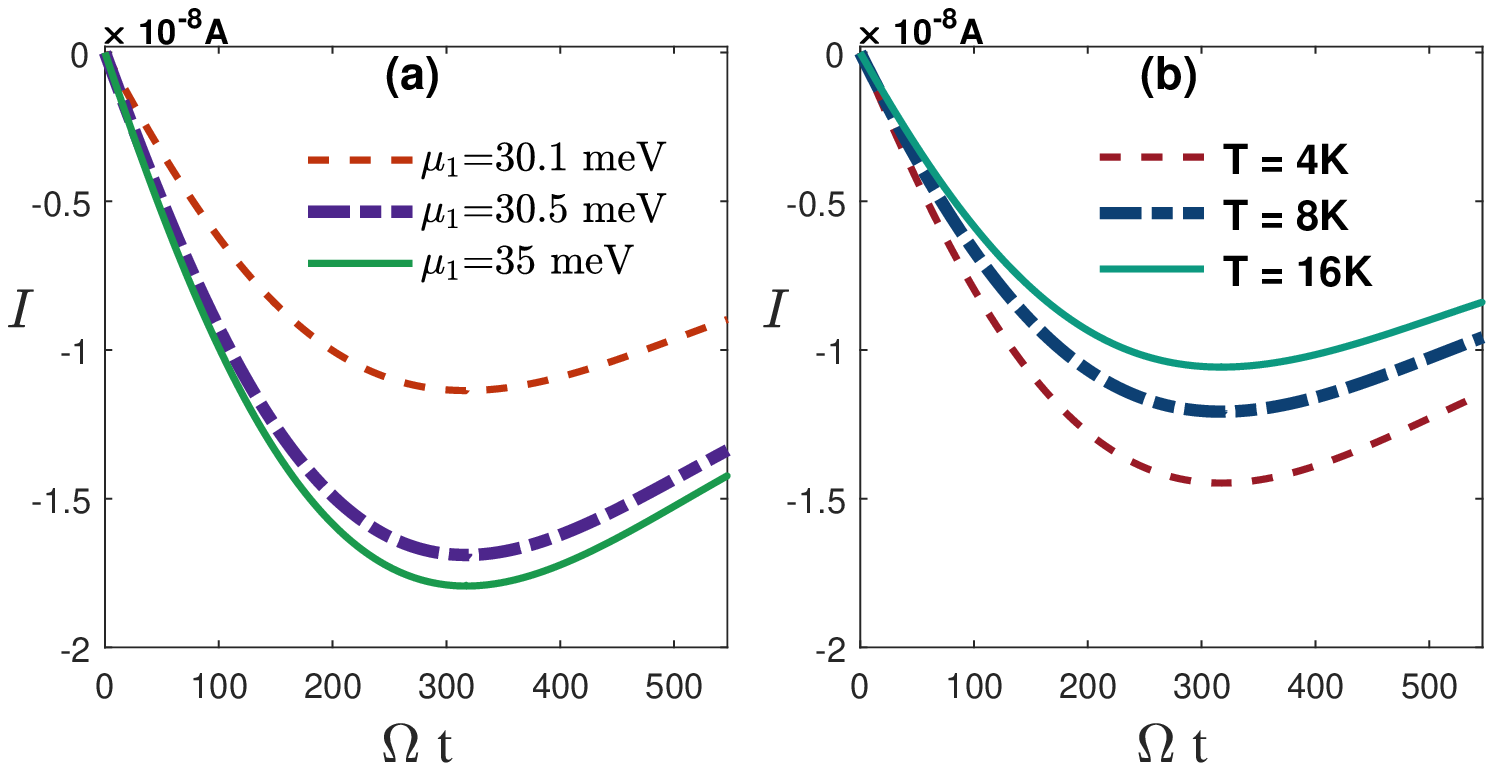}
 \caption{Single quantum dot interacting with double environments. The initial state of the system is:
 $\hrr(0) =|\rv_\rs\ra\la\rv_\rs|$. The commonly shared parameters are set as $\Lam=0.05$, $\vDe\om =2\pi {\rm Grad/s}$,
 $b=0.002$, $\Om=30$${\rm meV}$, $\mu_2=0$. The other parameters are set as
 (a) $T=2{\rm K}$; (b) $\mu_1=30.5$${\rm meV}$. } 
 \label{SingleDotDouEnvCurr3}
\end{figure}  
\begin{figure}[t]
 \centering\includegraphics[scale=0.57]{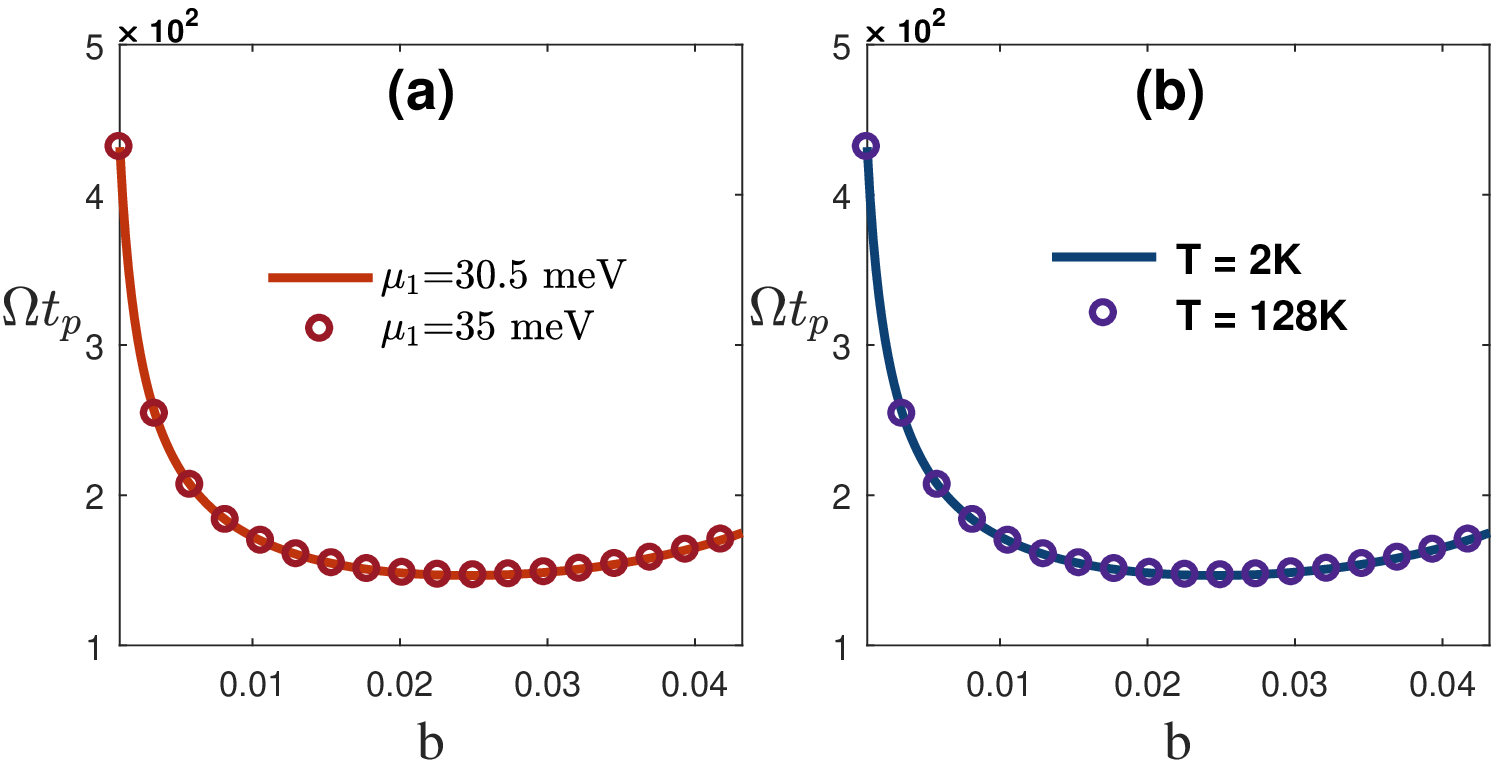}
 \caption{Single quantum dot interacting with double environments. The initial state of the system is:
 $\hrr(0) =|\rv_\rs\ra\la\rv_\rs|$. The commonly shared parameters are set as $\Lam=0.05$, $\vDe\om =2\pi {\rm Grad/s}$,
  $\Om=30$${\rm meV}$, $\mu_2=0$. The other parameters are set as
 (a) $T=2{\rm K}$; (b) $\mu_1=31$${\rm meV}$. } 
 \label{SingleDotDouEnvStPt}
\end{figure} 
becomes bigger, the current $I$ approaches the steady state faster, the oscillation of the current gets weaker, and beyond
a threshold of the bandwidth the oscillation disappears.
To evaluate how fast the current $I$ approaches its steady state in the extreme non-Markovian case,
we plot the stationary point of the current (where the time derivative equals 0) in Fig.~\ref{SingleDotDouEnvCurr3}.

\begin{figure}[!h]
 \centering\includegraphics[scale=0.6]{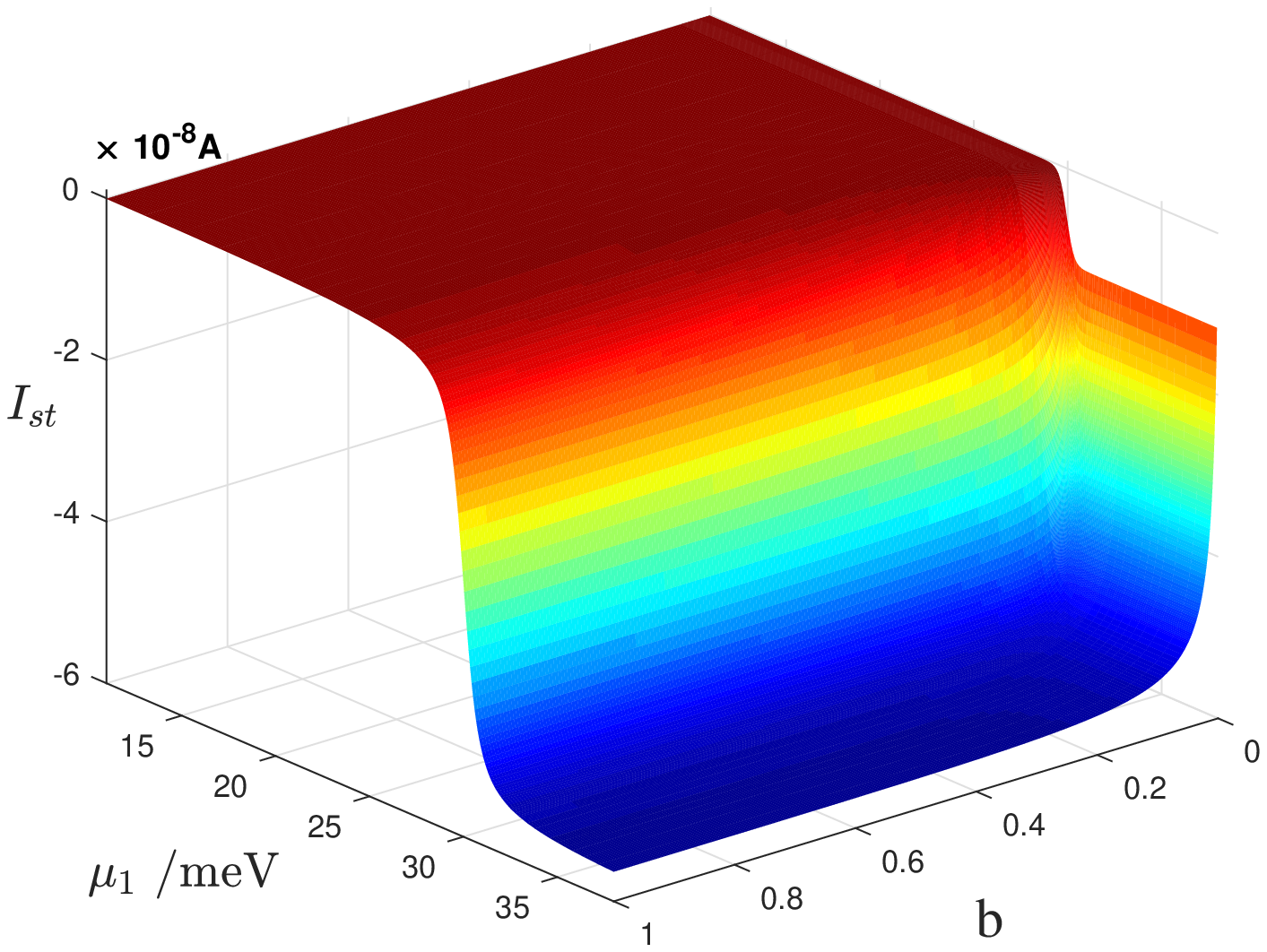}
 \caption{Single dot interacting with double environments. The initial state of the system is:
 $\hrr(0) =|\rv_\rs\ra\la\rv_\rs|$. The parameters are set as $\Lam=0.05$, $\vDe\om =2\pi {\rm Grad/s}$,
 $\Om=30$${\rm meV}$, $T=2{\rm K}$, $\mu_2=0$.} 
 \label{SingleDotDouEnvSteadyCurrent01}
\end{figure} 
Fig.~\ref{SingleDotDouEnvCurr3} shows that the chemical potential and temperature have no visible influence on the value $t_p$, where $t_p$ is the time coordinate corresponding to the first turning point of the current. Therefore, it shows that the bandwidth $b$ is the only important factor that can help $I$ to reach its steady value faster. In Fig.~\ref{SingleDotDouEnvStPt}, the irrelevances between $\mu_1$, $T$ and $t_p$ are further demonstrated
and we can see how the bandwidth influences the $t_p$. In both (a) and (b), the bandwidth range starts from $0.001$. 

Now we use ME to discuss the steady state current denoted as $I_{st} \eq I(t\!\raw\! \ift)$.
In Fig.~\ref{SingleDotDouEnvSteadyCurrent01}, we plot the 3-D figure of $I_{st}(b,\mu_1)$, and we see
that there are two relative flat regions corresponding to the top part and bottom part respectively,
and between these two there are two relative linear regions. Note that the bandwidth $b$ starts from $0.002$ in Fig.~\ref{SingleDotDouEnvSteadyCurrent01}.

\section{Effect of the Coulomb blockade in non-Markovian environments}
\label{Coulomb}

We now turn to the problem of the transport involving the Coulomb blockade effect. When there are more than one electrons in the dot or
there are more than one dots in the system, the Coulomb repulsion between electrons or dots may blockade the 
current~\cite{Gurvitz,you1999,Yan1,Yan2}.
Such an effect can be applied to measure the state of the dot or control the current.
In this section, we will use approximate MEs to investigate the Coulomb blockade effect in the non-Markovian case.
We consider two separate cases. In the first case, the spin degrees of freedom will be taken into account,
the second case will address a non-degeneracy situation, which will lead to interesting correlated noises phenomena.
 
\subsection{Coulomb blockade in the spin degeneracy situation}

In this model we suppose all the energy levels in the environments or in the dot contain two modes which correspond to spin up and spin
down, respectively. We assume that when the electrons tunnel in or tunnel out of the dot, their spins do not flip, as illustrated in
Fig.~\ref{figcoulomb3}. In this case, the Hamiltonian can be written as
\begin{eqnarray}
 \hh_\rt &=& \sum_{\si =\da, \ua} \Om \hd_\si\+ \hd_\si +\Om_c \hd_\da\+ \hd_\da \hd_\ua\+ \hd_\ua 
     +\sum_{k,\si} \big( \hd_\si\+ ( t_{1k} \hb'_{1k\si}  \non \\
 && +t_{2k}\hb'_{2k\si}) +( t_{1k} \hb'_{1k\si}{}\3\+ +t_{2k}\hb'_{2k\si}{}\3\+) \hd_\si \big)
     \non \\
 && +\sum_{k,\si} \om_k (\hb'_{1k\si}{}\3\+\, \hb'_{1k\si} + \hb'_{2k\si}{}\3\+\, \hb'_{2k\si}),
\end{eqnarray}
where $\hh_\rs=\sum_{\si =\da, \ua} \Om \hd_\si\+ \hd_\si +\Om_c \hd_\da\+ \hd_\da \hd_\ua\+ \hd_\ua$ is the Hamiltonian of the quantum dot, and
$\Om_c$ is the the Coulomb repulsion energy between the electrons.

\begin{figure}[h]
\begin{center}
\includegraphics[width=2.6in]{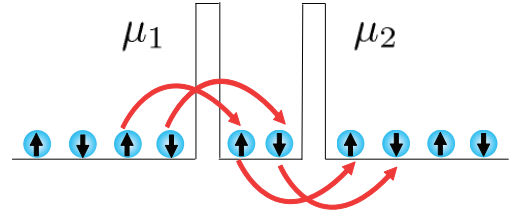}
\end{center}
\caption{A schematic diagram of a quantum dot coupled to two environments with
chemical potentials $\mu_{1}$ and $\mu_{2}$. A single electron either spin up or
spin down can be transported through the quantum dot.}
\label{figcoulomb3}
\end{figure}

The method used here closely follows the presentation in Sec.~\ref{Transport}.
By applying the purification and Bogoliubov transformations to the two environments in thermal equilibrium states,
we introduce eight independent Grassmann Gaussian stochastic processes:
\begin{eqnarray}
 \xi^*_{b_{j\si}t} &\eq& -i\sum_k \sq{1-\bar{n}_{jk\si}}\, t_{jk} e^{i\om_k t} \xi^*_{b_{jk\si}}  \non\\
 &\eq& -i\sum_k t_{b_{jk}} e^{i\om_k t} \xi^*_{b_{jk\si}},  \non \\
 \xi^*_{c_{j\si}t} &\eq& -i\sum_k \sq{\bar{n}_{jk\si}}\, t_{jk} e^{-i\om_k t} \xi^*_{c_{jk\si}}  \non\\
 &\eq& -i\sum_k t_{c_{jk}} e^{-i\om_k t} \xi^*_{c_{jk\si}},
\end{eqnarray}
 and altogether sixteen Q-operators corresponding to these Grassmann stochastic processes:
\begin {eqnarray}
 && \bq_\nu(t,\bx^*) \equiv \int_0^t ds K_\nu(t,s) \hq_\nu(t,s,\bx^*), \non\\
 && \hq_\nu(t,s,\bx^*) |\psi_t\ra \equiv \ora{\de}_{\xi^*_{\nu s}} |\psi_t(\bx^*)\ra,
\end{eqnarray}
where the non-commutative noises satisfy $\mm(\xi^*_{\nu t})=\mm(\xi^*_{\nu t}\xi^*_{\mu' s})=0$, and 
$\mm(\xi_{\nu t}\xi^*_{\nu' s})=\de_{\nu\nu'} K_\nu(t,s)$. Here $K_\nu(t,s)$ are the correlation functions of the noises, 
and the index $\nu=b_{1\da}, b_{1\ua},
 b_{2\da}, b_{2\ua}, c_{1\da}, c_{1\ua}, c_{2\da}, c_{2\ua}$. Since the spectrum is symmetric with respect to spin interchange, 
the correlation functions satisfy $K_{\lam_{j\da}}(t,s) =K_{\lam_{j\ua}}(t,s)$, where $\lam=b,c$ and $j=1,2$.
Thus, we will eliminate the spin index in the correlation functions in this subsection hereinafter:
\begin {eqnarray}
 K_{b_j}(t,s) &=& \sum_k t^2_{b_{jk}} e^{-i\om_k(t-s)},  \non \\
 K_{c_j}(t,s) &=& \sum_k t^2_{c_{jk}} e^{i\om_k(t-s)},  
\end {eqnarray}
Then the fermionic QSD equation can be derived:
\begin {eqnarray}
 \pa_t|\psi_t(\bx^*)\ra &=& \big( -i\hh_\rs +\sum_\si\!\! \sum_{j=1,2} (\xi^*_{b_{j\si} t} \hd_\si -\xi^*_{c_{j\si} t} \hd_\si\+  \non \\
 && -\hd_\si\+\bq_{b_{j\si}} +\hd_\si\bq_{c_{j\si}} )\big) |\psi_t(\bx^*)\ra.  \label{Coulomb1QSD1}
\end {eqnarray}
We can see that the functional form of the coupling terms with  $\xi^*_{\lam_{1\si}t}$ in Eq.~(\ref{Coulomb1QSD1}) is the same 
as the ones with $\xi^*_{\lam_{2\si}t}$, so
their functional derivatives should equal: 
$\ora{\de}_{\xi^*_{\lam_{1\si}s}}|\psi_t(\bx^*)\ra=\ora{\de}_{\xi^*_{\lam_{2\si}s}}|\psi_t(\bx^*)\ra$, which
means $\hq_{\lam_{1\si}}=\hq_{\lam_{2\si}}$. Thus, we will eliminate the index $j$ in $\hq_{\lam_{j\si}}$ in this subsection
hereinafter. $\bq_\nu$ can also be simplified. We define $K_{\lam}(t,s)\eq\sum_{j}K_{\lam_j}(t,s)$, and 
$\bq_{\lam_\si}\eq\sum_j\bq_{\lam_{j\si}}$, then we have $\bq_{\lam_\si}(t,\bx^*)=\int_0^t ds K_\lam(t,s)\hq_{\lam_{\si}}(t,s,\bx^*)$.

The evolution of the Q-operators is governed by the equations and boundary conditions:
\begin {eqnarray}
 \pa_t \hq_{\lam_\si}(t,s,\bx^*) &=& [-i\hh_\rs +\sum_\si\!\! \sum_{j=1,2} (\xi^*_{b_{j\si} t} \hd_\si -\xi^*_{c_{j\si} t} \hd_\si\+)  \non \\
 && +\sum_\si(\hd_\si\bq_{c_{\si}} -\hd_\si\+\bq_{b_{\si}}),  \hq_{\lam_\si}(t,s,\bx^*)]  \non \\
 && +\ora{\de}_{\!\xi^*_{\lam_{1\si} s}} \sum_\si(\hd_\si\bq_{c_{\si}} -\hd_\si\+\bq_{b_{\si}}),  \label{Qdynamics} \\
 &&\6\6\6\6\6\3\!\! \hq_{b_\si}(t\!=\!s,s,\bx^*)\!=\!\hd_{\si}, \q  \hq_{c_\si}(t\!=\!s,s,\bx^*)\!=\!-\hd\+_{\si},  \label{QdynamicsBound}
\end {eqnarray}
which can be derived from the consistency conditions: 
$\pa_t\ora{\de}_{\!\xi^*_{\lam_{j\si} s}}|\psi_t(\bx^*)\ra =\ora{\de}_{\!\xi^*_{\lam_{j\si} s}}\pa_t|\psi_t(\bx^*)\ra$.
In this model, the Q-operators will involve the noise terms up to infinite degree, however when the coupling strength between the
system and environments is weak,  the zeroth order approximation is proved to be a good approximation for perturbation.

If we neglect all the noise terms in Eq.~(\ref{Qdynamics}), an approximate equation can be obtained:
\begin {eqnarray}
 \pa_t \hq_{\lam_\si} \app [-i\hh_\rs +\sum_\si(\hd_\si\bq_{c_{\si}} -\hd_\si\+\bq_{b_{\si}}),  \hq_{\lam_\si}].  \q \label{Q0dynamics}
\end {eqnarray}
We use $\hq_{\lam_\si}\ze$ and $\bq_{\lam_\si}\ze$ to denote these approximate Q-operators.
Then Eq.~(\ref{Q0dynamics}) can be rewritten as
\begin {eqnarray}
 \pa_t \hq_{\lam_\si}\ze &=& [-i\hh_\rs +\sum_\si (-\hd_\si\+\bq_{b_{\si}}\ze +\hd_\si\bq_{c_{\si}}\ze), \hq_{\lam_\si}\ze], \q 
     \label{Q0dynamicsCou1}
\end {eqnarray}
where the boundary conditions are $\hq_{b_\si}\ze(t=s,s)=\hd_{\si}$ and $\hq_{c_\si}\ze(t=s,s)=-\hd_{\si}\+$.
When the Coulomb interaction is present, the zeroth order Q-operators are of such forms:
\begin{eqnarray}
 \bq_{b_{j\da}}\ze &=& F_{b_{j\da}}(t) \hd_\da +G_{b_{j\da}}(t)\hd_\da\hd_\ua\+\hd_\ua   \non\\
 &=& \int_0^t ds K_{b_{j}}(t,s) \Big(f_{b_{\da}}(t,s) \hd_\da +g_{b_{\da}}(t,s) \hd_\da\hd_\ua\+\hd_\ua \Big),   \non\\
 \bq_{c_{j\da}}\ze &=& F_{c_{j\da}}(t) \hd_\da\+ +G_{c_{j\da}}(t)\hd_\da\+\hd_\ua\+\hd_\ua   \non\\
 &=& \int_0^t ds K_{c_{j}}(t,s) \Big(f_{c_{\da}}(t,s) \hd_\da\+ +g_{c_{\da}}(t,s) \hd_\da\+\hd_\ua\+\hd_\ua \Big), \non\\
 \bq_{b_{j\ua}}\ze &=& F_{b_{j\ua}}(t) \hd_\ua +G_{b_{j\ua}}(t)\hd_\ua\hd_\da\+\hd_\da  \non \\
 &=& \int_0^t ds K_{b_{j}}(t,s) \Big(f_{b_{\ua}}(t,s) \hd_\ua +g_{b_{\ua}}(t,s) \hd_\ua\hd_\da\+\hd_\da \Big),  \non \\
 \bq_{c_{j\ua}}\ze &=& F_{c_{j\ua}}(t) \hd_\ua\+ +G_{c_{j\ua}}(t)\hd_\ua\+\hd_\da\+\hd_\da  \non \\
 &=& \int_0^t ds K_{c_{j}}(t,s) \Big(f_{c_{\ua}}(t,s) \hd_\ua\+ +g_{c_{\ua}}(t,s) \hd_\ua\+\hd_\da\+\hd_\da \Big). \q\q \label{Cou1Expansion}
\end{eqnarray}
Due to the symmetry with respect to spin interchange in Eq.~(\ref{Q0dynamicsCou1}), we have the following relations:
$f_{\lam_\da}=f_{\lam_\ua}$, $g_{\lam_\da}=g_{\lam_\ua}$, $F_{\lam_{j\da}}=F_{\lam_{j\ua}}$, $G_{\lam_{j\da}}=G_{\lam_{j\ua}}$,
$\lam=b,c$, and $j=1,2$. Subsequently, we may drop the index $\si$ in $f_{\lam_\si}$, $g_{\lam_\si}$, $F_{\lam_{j\si}}$ and $G_{\lam_{j\si}}$ in
this subsection from now on. By substituting the expansions in Eq.~(\ref{Cou1Expansion}) into Eq.~(\ref{Q0dynamicsCou1}),
the equations of $f_\lam$ and $g_{\lam}$ can be obtained:
\begin{eqnarray}
 \pa_t f_b(t,s) &=& \big(i\Om +\!\sum_\lam \!\!\sum_{j=1,2}\!\!F_{\lam_j}(t) -\sum_{j=1,2}\!\!G_{c_j}(t) \big) f_{b}(t,s),   \non \\
 \pa_t f_c(t,s)  &=& -\big(i\Om +\!\sum_\lam \!\!\sum_{j=1,2}\!\!F_{\lam_j}(t) -\sum_{j=1,2}\!\!G_{c_j}(t) \big) f_{c}(t,s),   \non \\
 \pa_t g_b(t,s) &=& \big(i\Om + \!\sum_\lam \!\!\sum_{j=1,2} F_{\lam_j}(t) -\sum_{j=1,2}\!\!G_{c_j}(t) \big) g_{b}(t,s)   \non \\
 && +\big(i\Om_c +2\!\sum_\lam \!\!\sum_{j=1,2} G_{\lam_j}(t) \big) \big( f_b(t,s) +g_b(t,s) \big), \non \\
 \pa_t g_c(t,s) &=& -\big(i\Om +\!\sum_\lam \!\!\sum_{j=1,2} F_{\lam_j}(t) -\sum_{j=1,}\!\!G_{c_j}(t) \big) g_{c}(t,s)   \non \\
 && -\big(i\Om_c +2\!\sum_\lam \!\!\sum_{j=1,2} G_{\lam_j}(t) \big) \big( f_c(t,s) +g_c(t,s) \big),  \non \\
\end{eqnarray}
where the boundary conditions are $f_b(t\!=\!s,s)\!=\!1$, $f_e(t\!=\!s,s)\!=\!-1$, $g_\lam(t\!=\!s,s)\!=\!0$.

Once the coefficients of Q-operators are solved,
we then can easily derive the ME for the density operator of the system:
\begin{eqnarray}
 \pa_t \hrr &=& -i\left[\Om\hd_\da\+\hd_\da +\Om\hd_\ua\+\hd_\ua +\Om_c\hd_\da\+\hd_\da\hd_\ua\+\hd_\ua , \hrr\right]  \non\\
 && +\Big( \sum_{\si} (F_c\hd_\si\hd_\si\+ -F_b\hd_\si\+\hd_\si +G_c\hd_\si\+\hd_\si)\hrr   \non\\
 && -2(G_b +G_c) \hd_\da\+\hd_\da\hd_\ua\+\hd_\ua  \big) \hrr 
     +G_b \hd_\da\hd_\ua\+\hd_\ua \hrr \hd_\da\+  \non \\
 && -G_c \hd_\da\+\hd_\ua\+\hd_\ua \hrr \hd_\da +G_b \hd_\ua\hd_\da\+\hd_\da \hrr \hd_\ua\+
     -G_c \hd_\ua\+\hd_\da\+\hd_\da \hrr \hd_\ua  \non \\
 && +\rhc \Big) +2\sum_{\si} (F_b^R \hd_\si \hrr \hd_\si\+ -F_c^R \hd_\si\+\hrr\hd_\si),
\end {eqnarray}
where $F_\lam \eq \sum_{j=1,2}F_{\lam_j}$, and $G_\lam \eq \sum_{j=1,2} G_{\lam_j}$; the superscript $R$ stands for the real part of 
a complex number.

Following the same procedure introduced in Sec.~\ref{Transport}, we can find the electronic current
between the quantum dot and the $1$st environment:
\begin{eqnarray}
 I_{1d} \!&=& q_e\!\Tr\Big(\!\big(\! -\!2 \sum_\si (G_{c_1}^R \hd_\si\+\hd_\si \!+\!F_{c_1}^R \hd_\si \hd_\si\+) 
     \!-\! 2F_{b_1}^R (\sum_\si \hd_\si\+\hd_\si)^2  \non \\
 && \!+4(F_{b_1}^R \!-\!G_{b_1}^R \!+\!G_{c_1}^R) \hd_\da\+\hd_\da\hd_\ua\+\hd_\ua 
     \big)  \hrr \!\Big),  \non \\
 &=& -2q_e \big( 2F_{c_1}^R \rho_{00} +(F_{c_1}^R +G_{c_1}^R +F_{b_1}^R)(\rho_{\da\da} +\rho_{\ua\ua})  \non \\
 && +2(F_{b_1}^R +G_{b_1}^R)\rho_{22} \big),  \q \label{t44}
\end{eqnarray}
where we expand the reduced density operator as
\begin {eqnarray}
 \hrr &=& \rho_{00} |0\ra \la0| +\rho_{\da\da} |\e\da\ra \la\da\e| +\rho_{\ua\ua}|\e\ua\ra \la\ua\e| 
     +\rho_{22} |\e\ua\ra|\e\da\ra \la\da\e|\la\ua\e|  \non \\
 && +\rho_{\da\ua}|\e\da\ra \la\ua\e| +\rho_{\da\ua}^*|\e\ua\ra \la\da\e|.  \label{t45}
\end {eqnarray}
The current $I_{d2}$ can be obtained similarly:
\begin {eqnarray}
 I_{d2} &=& 2q_e \big(2F_{c_2}^R \rho_{00} +(F_{c_2}^R +G_{c_2}^R +F_{b_2}^R)(\rho_{\da\da} +\rho_{\ua\ua})  \non \\
 && +2(F_{b_2}^R +G_{b_2}^R)\rho_{22} \big).
\end {eqnarray}

In the numerical simulation part, we still use the Lorentzian spectrum:
\begin {eqnarray}
 t_{jk}^2 =\Lam^2\Om\vDe\om \fr{b^2}{(\om_k/\Om -1)^2 +b^2},  \q (j=1,2).
\end {eqnarray}
where for simplicity the couplings to the source and drain are assumed to be symmetric.

\begin{figure}[htb]
\begin{center}
\includegraphics[scale=0.58]{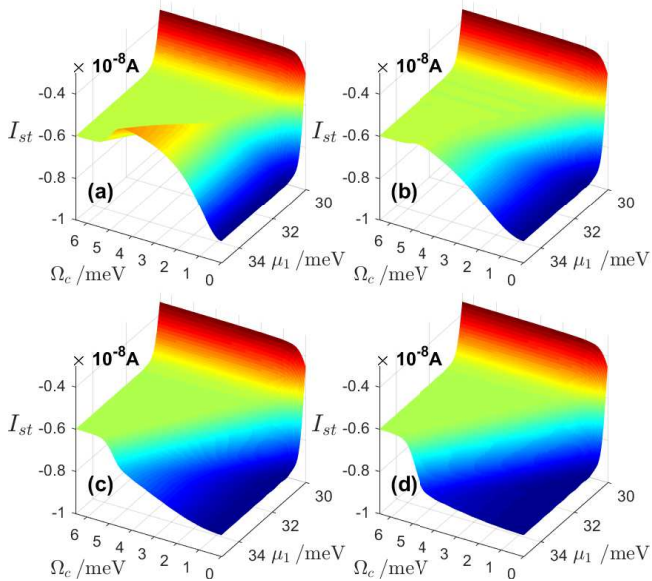}
\end{center}
\caption{The currents $I_{st}$ evaluated from the approximate
zeroth order ME, and plotted as a function of the chemical potential $\mu_1$ and the energy of Coulomb interaction
$\Omega_c$. The parameters are set as $\vDe\om=2\pi$Grad/s, $T=2$K,  $\mu_2= 0$, $\Om=30$ meV, 
and $\Lam=0.014$. The bandwidths are respectively set as (a) $b=0.05$; (b) $b=0.1$; (c) $b=0.2$; (d) $b=0.4$. }
\label{Coulomb1current1}
\end{figure}

\begin{figure}[!htb]
\begin{center}
\includegraphics[scale=0.64]{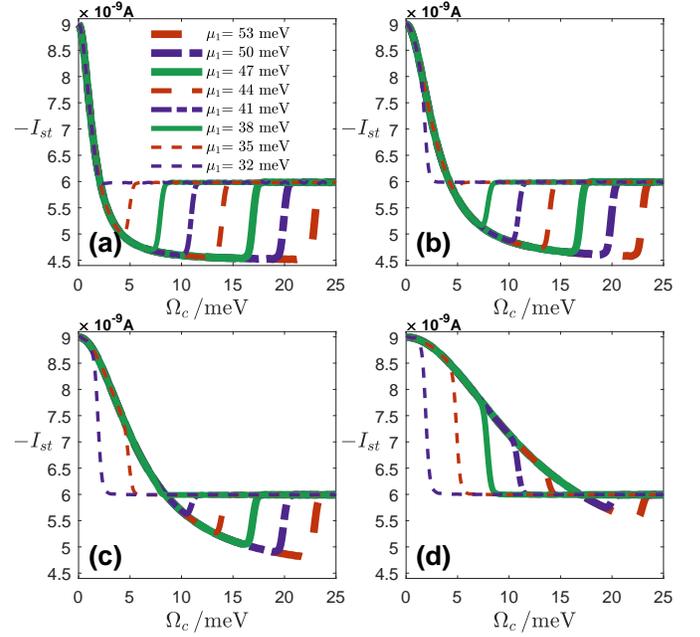}
\end{center}
\caption{The currents $I_{st}$ evaluated from the approximate
zeroth order ME, and plotted as a function of the energy of Coulomb interaction $\Omega_c$. 
The commonly shared parameters are set as $\vDe\om=2\pi$Grad/s, $T=2$K,  $\mu_2= 0$, $\Om=30$ meV, and $\Lam=0.014$.
The bandwidths are respectively set as (a) $b=0.05$; (b) $b=0.1$; (c) $b=0.2$; (d) $b=0.4$.}
\label{Coulomb1current2}
\end{figure}
To make the approximation valid, we set $\Lam=0.014$ and $b\geq 0.05$ in this subsection. In this model the steady state of $\hrr$ can be 
fully determined by the coefficients $F_\lam$ and $G_\lam$, thus we do not indicate the initial state of the system. 
From Fig.~\ref{Coulomb1current1}, we can see how the Coulomb energy $\Om_c$ influences the steady state current 
$I_{st}\eq(I_{1d}(t\!\raw\!\ift) +I_{d2}(t\!\raw\!\ift))/2$. In all the four subfigures which represent different
bandwidths, there is a clear line corresponding to $\mu_1=\Om+\Om_c$. In the region of $\mu_1\leq\Om+\Om_c$, the steady state current
keeps almost a constant value which is just the blockaded current. Fig.~\ref{Coulomb1current2} displays some sections of 
Fig.~\ref{Coulomb1current1} perpendicular to axis $\mu_1$.
We can see that,  when the Coulomb interaction is ignored, the steady state current is given by $-$9\,nA, when $\Om_c \raw \ift$, the steady state current  converges to about $2/3$ of $-$9\,nA. Before reaching $2/3$ of the current $I_{st}(\Om_c=0)$, the steady state current may have higher values, which are dependent on $\mu_1$. When $\mu_1$ increases, the overshoot will converge
to $1/2$ of the current $I_{st}(\Om_c=0)$.

\subsection{Coulomb blockade in the correlated noise situation}

Suppose two quantum dots are arranged in parallel between the source and drain, and the spin degrees of freedom of all electrons
are neglected, (the non-degeneracy case can be achieved by adding a strong magnetic field) then for each electron in the environments,
it has a chance to tunnel into either dot. A schematic configuration for this situation is shown in Fig.~\ref{figcoulomb4}, as to be shown below,
this case will generate correlated noises.

\begin{figure}[h]
\begin{center}
\includegraphics[scale=0.62]{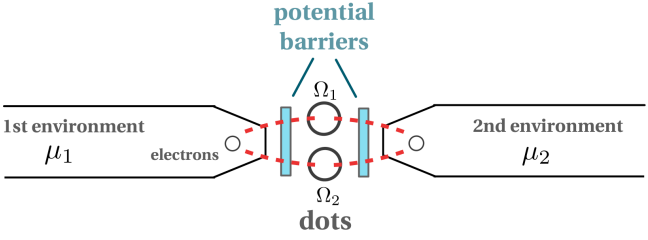}
\end{center}
\caption{A schematic diagram of a single quantum dot with two electrons
transported between two reservoirs, each mode of the reservoirs are coupled to
both electrons simultaneously.}
\label{figcoulomb4}
\end{figure}

The Hamiltonian of the total system may be written as
\begin {eqnarray}
 \hh_\rt &=& \!\sum_{\ep=1,2}\Om_\ep\hd_\ep\+\hd_\ep +\Om_c \hd_1\+\hd_1\hd_2\+\hd_2    
     +\sum_{k,\ep} \big(\hd_\ep\+( t_{\ep,1k}\hb'_{1k}  \non \\
 && +t_{\ep,2k} \hb'_{2k}) +(t_{\ep,1k} \hb'_{1k}{}\e\+ +t_{\ep,2k}\hb'_{2k}{}\e\+) \hd_\ep \big)   \non \\
 && +\sum_k \om_k (\hb'_{1k}{}\e\+\hb'_{1k} +\hb'_{2k}{}\e\+\hb'_{2k}),   
\end {eqnarray}
where $\hh_\rs=\sum_{\ep=1,2}\Om_\ep\hd_\ep\+\hd_\ep +\Om_c \hd_1\+\hd_1\hd_2\+\hd_2$ is the Hamiltonian of the dots, and 
$\Om_c$ is the Coulomb repulsive energy. Again, the parameter $t_{\ep,jk}$ is taken as a real number representing the coupling between the $\ep$th dot to the $k$th mode of the $j$th environment ($j=1,2$).

A set of Grassmann stochastic processes can be obtained by applying the purification and Bogoliubov transformations to the environments, 
\begin {eqnarray}
 \xi^*_{\ep b_j t} &\eq& -i\sum_k \sq{1-\bar{n}_{jk}}\, t_{\ep,jk} e^{i\om_k t}\xi^*_{b_{jk}}  \non \\
 &\eq& -i\sum_k t_{\ep b_{jk}} e^{i\om_k t} \xi^*_{b_{jk}},  \non \\
 \xi^*_{\ep c_j t} &\eq& -i\sum_k \sq{\bar{n}_{jk}}\, t_{\ep,jk} e^{-i\om_k t}\xi^*_{c_{jk}}  \non \\
 &\eq& -i\sum_k t_{\ep c_{jk}} e^{-i\om_k t} \xi^*_{c_{jk}},
\end {eqnarray}
which satisfy $\mm(\xi^*_{\ep\lam_jt})=\mm(\xi^*_{\ep\lam_jt}\xi^*_{\ep'\lam'_{j'}s})=0$. 
The noise $\xi^*_{\ep\lam_jt}$ represents the stochastic influence of the environment $\lam_j$ on the dot $\ep$.
Usually $\xi^*_{\ep\lam_jt}$ and $\xi^*_{\ep'\lam_js}$ are highly correlated, the elements of the cross correlation function matrix is defined 
as
\begin {eqnarray}
 C_{\ep\ep',\nu}(t,s) \equiv \mm \big(\xi_{\ep\nu t} \xi_{\ep'\nu s}^* \big).
\end {eqnarray}
where the subscript $\nu$ is a short notation of $\lam_j$ which represents the four possible indices $b_1$, $b_2$, $c_1$, $c_2$.
 
Then the equation of motion for the fermionic quantum trajectory can be obtained:
\begin {eqnarray}
 && \pa_t|\psi_t(\bx^*)\ra  \non \\
 &=& \big(\!-\!i\hh_\rs \!+\!\sum_{\ep,j} \hd\+_\ep (\xi^*_{\ep c_j t} 
     -\e\int_0^t \!ds \sum_{\ep'} \!C_{\ep\ep'\!,b_j}\!(t,s) \!\ora{\de}_{\e\xi^*_{\ep'b_j\!s}})  \non \\
 && \e+\!\sum_{\ep,j} \hd_\ep (-\xi^*_{\ep b_j t} \!+\e\int_0^t \!ds \sum_{\ep'} \!C_{\ep\ep'\!,c_j}\!(t,s) \!\ora{\de}_{\e\xi^*_{\ep'\!c_j\!s}} ) \!\big)
     \!|\psi_t(\bx^*)\ra.  \q\;\;\;  \label{Coulomb2QSD1}
\end {eqnarray}
The functional derivatives  in Eq.~(\ref{Coulomb2QSD1}) can be replaced by Q-operators:
\begin {eqnarray}
 &&\hq_{\ep\nu}(t,s,\bx^*)|\psi_t(\bx^*)\ra =\ora{\de}_{\xi^*_{\ep\nu s}}|\psi_t(\bx^*)\ra,  \non \\
 &&\bq_{\ep\nu}(t,\bx^*) \eq \int_0^t ds \sum_{\ep'} C_{\ep\ep',\nu}(t,s) \hq_{\ep'\nu}(t,s,\bx^*), \q
\end {eqnarray}
Then the fermionic QSD equation is finally obtained:
\begin {eqnarray}
 \pa_t|\psi_t(\bx^*)\ra &=& \big(-i\hh_\rs +\sum_{\ep,j} \hd\+_\ep (\xi^*_{\ep c_j t} -\bq_{\ep b_j})  \non \\
 \3&&\3 +\sum_{\ep,j} \hd_\ep(\bq_{\ep c_j} -\xi^*_{\ep b_j t}) \Big) |\psi_t(\bx^*)\ra.  \label{Coulomb2QSD2}
\end {eqnarray}

By a similar observation on the coupling forms of $\xi^*_{\ep\lam_1t}$ and $\xi^*_{\ep\lam_2t}$ in Eq.~(\ref{Coulomb2QSD2}),
we conclude that the Q-operators should satisfy: $\hq_{\ep\lam_1} =\hq_{\ep\lam_2}$. 
Thus, we may drop the index $j$ in $\hq_{\ep\lam_j}$ from now on in this subsection.

When the couping is weak, we may use $\hq_{\ep\lam}\ze$ to replace the exact Q-operators, where
the approximate operators evolve according yo the following equations:
\begin {eqnarray}
 \pa_t \hq\ze_{\ep\lam}(t,s) &=& [-i\hh_\rs+\sum_{\ep,j} (\hd_\ep\bq_{\ep c_j}\ze -\hd\+_\ep \bq_{\ep b_j}\ze), \hq^{\{0\}}_{\ep\lam}(t,s)],  \non \\
 && \6\6\6\6 \hq\ze_{\ep b}(t=s,s)=\hd_\ep, \q  \hq\ze_{\ep c}(t=s,s)=-\hd_\ep\+.    \label{Q0dynamicsCou2}
\end {eqnarray}
We expand $\bq_{\ep\lam_j}\ze$ and $\hq_{\ep\lam}\ze$ with respect to the annihilation and creation operators of the dots:
\begin {eqnarray}
 \bq_{\ep b_{j}}\ze &=& \sum_{\ep'=1,2} (F_{\ep\ep'b_{j}}(t) \hd_{\ep'} +G_{\ep\ep'b_{j}}(t) \hd_{\ep'}\hd_1\+\hd_1\hd_2\+\hd_2)   \non \\
 &\eq& \int_0^t ds \sum_{\ep'',\,\ep'} C_{\ep\ep'',b_{j}}(t,s) \Big( f_{\ep''\ep'b}(t,s)\hd_{\ep'}  \non \\
 && +g_{\ep''\ep'b}(t,s) \hd_{\ep'}\hd_1\+\hd_1\hd_2\+\hd_2 \Big),   \non \\
 \bq_{\ep c_{j}}\ze &=& \sum_{\ep'=1,2} (F_{\ep\ep'c_{j}}(t) \hd_{\ep'}\+ +G_{\ep\ep'c_{j}}(t) \hd_1\+\hd_1\hd_2\+\hd_2\hd_{\ep'}\+)   \non \\
 &\eq& \int_0^t ds \sum_{\ep'',\,\ep'} C_{\ep\ep'',c_{j}}(t,s) \Big( f_{\ep''\ep'c}(t,s)\hd_{\ep'}\+  \non \\ 
 && +g_{\ep''\ep'c}(t,s) \hd_1\+\hd_1\hd_2\+\hd_2\hd_{\ep'}\+ \Big).   
\end {eqnarray}
By substituting these expansions into Eq.~(\ref{Q0dynamicsCou2}), we get the equations and boundary conditions of the coefficients:
\begin {eqnarray}
 \pa_t \e
 \begin {bmatrix}
  f_{\ep1b} \\
  f_{\ep2b} \\
  g_{\ep1b} \\
  g_{\ep2b}
 \end {bmatrix} &=&
 \begin {bmatrix}
  A & D & 0 & 0 \\
  C & B & 0 & 0 \\
  E & 0 & A\!+\!E & D \\
  0 & E & C & B\!+\! E \\
 \end {bmatrix} \e
 \begin {bmatrix}
  f_{\ep1b} \\
  f_{\ep2b} \\
  g_{\ep1b} \\
  g_{\ep2b}
 \end {bmatrix},  \non \\
 \pa_t \e
 \begin {bmatrix}
  f_{\ep1c} \\
  f_{\ep2c} \\
  g_{\ep1c} \\
  g_{\ep2c}
 \end {bmatrix} &=&-
 \begin {bmatrix}
  A & C & 0 & 0 \\
  D & B & 0 & 0 \\
  E & 0 & A\!+\!E & C \\
  0 & E & D & B\!+\!E \\
 \end {bmatrix} \e
 \begin {bmatrix}
  f_{\ep1c} \\
  f_{\ep2c} \\
  g_{\ep1c} \\
  g_{\ep2c}
 \end {bmatrix},  \non \\     
 &&\6\6\6\6\6 f_{\ep\ep'b}(t\!=\!s,s) =\de_{\ep\ep'}, \q f_{\ep\ep'c}(t\!=\!s,s) =-\de_{\ep\ep'}, \non \\
 &&\6\6\6\6\6 g_{\ep\ep'\lam}(t\!=\!s,s) =0, \q  (\lam=b,c),  \label{Coulomb2f}
\end {eqnarray}
where the explicit expressions of $A$, $B$, $C$, $D$ and $E$ are
\begin {eqnarray}
 A &=& i\Om_1 +\sum_{\lam,j} F_{11\lam_j} -\sum_j G_{22c_j}, \non  \\
 B &=& i\Om_2 +\sum_{\lam,j} F_{22\lam_j} -\sum_j G_{11c_j},  \non \\
 C &=& \sum_j (F_{12b_j} +F_{21c_j} +G_{21c_j}), \non \\
 D &=& \sum_j (F_{21b_j} +F_{12c_j} +G_{12c_j}),  \non \\
 E &=& i\Om_c +\sum_{\ep,\lam,j} G_{\ep\ep\lam_j}.  \label{AB}
\end {eqnarray}

When the coefficients above are solved, we directly obtain the zeroth order ME through the extended Novikov theorem:
\begin{eqnarray}
 \pa_t \hrr &=& -i[\hh_\rs, \hrr] +\sum_{\ep,j} \big( [  \bq_{\ep b_j}\ze \hrr,\hd_\ep\+ ]  +[\hd_\ep, \bq_{\ep c_j}\ze \hrr ]  \non \\
 && +\rhc \big).
\end{eqnarray}
Then the current $I_{1d}$ and $I_{d2}$ can be separated from the ME:
\begin{eqnarray}
 I_{1d} &=& -2q_e \Big( \sum_\ep F_{\ep\ep c_1}^R \rho_{00} +(F_{11b_1}^R +F_{22c_1}^R +G_{22c_1}^R)\rho_{11}  \non \\
 && +(F_{11c_1}^R\!+\!F_{22b_1}^R\!+\!G_{11c_1}^R)\rho_{22}  \!+\!\sum_\ep (F_{\ep\ep b_1}^R \!+\!G_{\ep\ep b_1}^R )\rho_{33}  \non \\
 && +\big((F_{21b_1}\e+\e F^*_{12b_1}\e-\e F_{12c_1} \e-\e F^*_{21c_1}\e-\e G_{12c_1}\e-\e G^*_{21c_1})\rho_{12} \big)\!^R \Big),  \non \\
 I_{d2} &=& 2q_e \Big( \sum_\ep F_{\ep\ep c_2}^R \rho_{00} +(F_{11b_2}^R +F_{22c_2}^R +G_{22c_2}^R)\rho_{11}  \non \\
 && +(F_{11c_2}^R\!+\!F_{22b_2}^R\!+\!G_{11c_2}^R)\rho_{22}  +\sum_\ep (F_{\ep\ep b_2}^R \!+\!G_{\ep\ep b_2}^R )\rho_{33} \non \\
 && +\big( (F_{21b_2}\e+\e F^*_{12b_2}\e-\e F_{12c_2} \e-\e F^*_{21c_2} \e-\e G_{12c_2}\e-\e G^*_{21c_2}\!)\rho_{12} \big)\!^R \Big), 
     \non \\  \label{Coulomb2current1}
\end{eqnarray}
where the reduced density operator is expanded as
\begin{eqnarray}
 \hrr &=& \rho_{00}|0\ra\la0| +\rho_{11}\hd_1\+|0\ra\la0|\hd_1 +\rho_{22}\hd_2\+|0\ra\la0|\hd_2  \non \\
 &&\3 +\rho_{33}\hd_2\+\hd_1\+|0\ra\la0|\hd_1\hd_2 \!+\!\rho_{12}\hd_1\+|0\ra\la0|\hd_2 \!+\!\rho_{12}^*\hd_2\+|0\ra\la0|\hd_1.  \non \\
     \label{t50}
\end{eqnarray}

When the resonance condition is satisfied $\Om_1=\Om_2$,  it is easy to see that the steady state current will be affected by
the initial condition of $\hrr$, which is different from the model in the previous subsection.
In numerical simulations, we use such a Lorentzian spectrum form:
\begin {eqnarray}
 t_{\ep,jk}^2 \!=\!\sum_{\ep=1,2} \!\fr{\Om_\ep}{2} \!\cdot\! \fr{\Lam^2\vDe\om b^2}{(\om_k/\Om_\ep -1)^2 +b^2}, \q  (\ep,j\!=\!1,2).\q
\end {eqnarray}
With the chosen spectrum, when the resonance condition is valid ($\Om_1=\Om_2$), then all the four components in the cross correlation function are 
identical:
$C_{11,\lam_j}=C_{12,\lam_j}=C_{21,\lam_j}=C_{22,\lam_j}$. By substituting these relations into Eq.~(\ref{Coulomb2f}) and (\ref{AB}),
we get $f_{11\lam}=f_{12\lam}=f_{21\lam}=f_{22\lam}$, $g_{11\lam}=g_{12\lam}=g_{21\lam}=g_{22\lam}$, 
$F_{11\lam_j}=F_{12\lam_j}=F_{21\lam_j}=F_{22\lam_j}$ and $G_{11\lam_j}=G_{12\lam_j}=G_{21\lam_j}=G_{22\lam_j}$.
To simplify the notation, we drop the first two indices of the above coefficients.
With these symmetries, Eq.~(\ref{Coulomb2current1}) can be simplified to
\begin {eqnarray}
 I_{1d} &=& -4q_e \big(F_{c_1}^R \rho_{00} +(F_{b_1}^R +F_{c_1}^R +G_{c_1}^R)\fr{\rho_{11} +\rho_{22}}{2}  \non \\
 && +(F_{b_1}^R +G_{b_1}^R )\rho_{33}  +(F_{b_1}^R -F_{c_1}^R -G_{c_1}^R)\rho_{12}^R \big),  \non \\
 I_{d2} &=& 4q_e \big(F_{c_2}^R \rho_{00} +(F_{b_2}^R +F_{c_2}^R +G_{c_2}^R) \fr{\rho_{11} +\rho_{22}}{2}  \non \\
 && +(F_{b_2}^R +G_{b_2}^R )\rho_{33}  +(F_{b_2}^R -F_{c_2}^R -G_{c_2}^R)\rho_{12}^R \big). \q  \label{Coulomb2current2}
\end {eqnarray}

Due to the restriction on fermionic density operators, we only need to consider some of the matrix elements, which can be rearranged into a row vector, then the ME may be written as
\begin {footnotesize}
\begin {eqnarray}
 \pa_t \e
 \begin {bmatrix}
  \rho_{00} & \rho_{11} & \rho_{22} & \rho_{33} & \rho_{12}^R & \rho_{12}^I
 \end {bmatrix} \!=\!
 \begin {bmatrix}
  \rho_{00} & \rho_{11} & \rho_{22} & \rho_{33} & \rho_{12}^R & \rho_{12}^I
 \end {bmatrix}\e M^T\e,  \non \\ \label{Coulomb2ME1}
\end {eqnarray}
\end {footnotesize}
where the superscript $T$ denotes the transpose of a matrix, and $M$-matrix is of the form:
\begin {footnotesize}
\begin {eqnarray}
 M =
 \begin {bmatrix}
  2A   & B     & B     & 0    & 2B       & 0        \\
  -A  & C\!-\!B  & 0      & F   & -C\!-\!B   & -D\!-\!E   \\
  -A  & 0      & C\!-\!B  & F   & -C\!-\!B   & D\!+\!E    \\
  0    & -C    & -C    & -2F  & 2C       & 0        \\
  -A  & -(\!C\e+\e B)\!/2   & -(\!C\e+\e B)\!/2   & -F  & C\!-\!B    & 0        \\
  0    & (\!D\e+\e E)\!/2    & -(\!D\e+\e E)\!/2   & 0    & 0        & C\!-\!B    
 \end {bmatrix}\e.  \q\q \label{M}
\end {eqnarray}
\end {footnotesize}
The explicit expressions of the coefficients in $M$ are
\begin{eqnarray}
 A &\eq& 2(F_{c_1}^R +F_{c_2}^R), \q B \eq 2(F_{b_1}^R +F_{b_2}^R),  \non \\
 C &\eq& 2(F_{c_1}^R +G_{c_1}^R +F_{c_2}^R +G_{c_2}^R),  \non \\
 D &\eq& 2(F_{c_1}^I +G_{c_1}^I +F_{c_2}^I +G_{c_2}^I),  \non \\
 F &\eq& 2(F_{b_1}^R +G_{b_1}^R +F_{b_2}^R +G_{b_2}^R),  \q E\eq 2(F_{b_1}^I +F_{b_2}^I), \q\q
\end{eqnarray}
where the superscript $I$ stands for the imaginary part of a complex number.

The rank of $M$ contains a wealth of information about the steady-state Coulomb blockade. $M$ is generally a rank four matrix, however, the rank 
can be five when the resonance condition is satisfied, which reduces to the degeneracy case considered in the previous subsection. 
Therefore, we will be focused on the rank four case where the steady state current $I_{st}\!\eq\! (I_{1d}(t\!\raw\!\ift) +I_{d2}(t\!\raw\!\ift))/2$ is shown to be sensitively dependent on the initial state $\hrr(0)$.  Clearly, by the linearity of MEs, we only to consider the steady states of a complete basis of $\hrr(0)$, which can generate an arbitrary initial $\hrr(0)$ by a convex combination. For a reason to become clear later, we select following eight specific initial states as the basis ($i= 1, ..., 8$):
\begin {small}
\begin {eqnarray}
 \begin {bmatrix}
  \rho_{00,i}(0) \\
  \rho_{11,i}(0) \\
  \rho_{22,i}(0) \\
  \rho_{33,i}(0) \\
  \rho_{12,i}^R(0) \\
  \rho_{12,i}^I(0) 
 \end {bmatrix}
 \!=\!
 \begin {bmatrix}
  1 \\
  0 \\
  0 \\
  0 \\
  0 \\
  0 
 \end {bmatrix}\!,\!
 \begin {bmatrix}
  0 \\
  0.5 \\
  0.5 \\
  0 \\
  0.5 \\
  0
 \end {bmatrix}\!,\!
 \begin {bmatrix}
  0 \\
  1 \\
  0 \\
  0 \\
  0 \\
  0
 \end {bmatrix}\!,\!
 \begin {bmatrix}
  0 \\
  0 \\
  1 \\
  0 \\
  0 \\
  0
 \end {bmatrix}\!,\!
 \begin {bmatrix}
  0 \\
  0.5 \\
  0.5 \\
  0 \\
  0 \\
  0.5
 \end {bmatrix}\!,\!
 \begin {bmatrix}
  0 \\
  0.5 \\
  0.5 \\
  0 \\
  0 \\
  -0.5
 \end {bmatrix}\!,\!
 \begin {bmatrix}
  0 \\
  0 \\
  0 \\
  1 \\
  0 \\
  0
 \end {bmatrix}\!,\!
 \begin {bmatrix}
  0 \\
  0.5 \\
  0.5 \\
  0 \\
  -0.5 \\
  0
 \end {bmatrix}\!,\! \non \\ \label{CouIni}
\end {eqnarray}
\end {small}
Obviously, the choice of a basis is not unique. In Fig.~\ref{CouIni01}, we show that the eight initial basis states may be divided into three groups.
The group I consists of first two initial vectors in Eq.~(55), Group II contains four initial states ranging from 3 to 6; Group III contains the rest two initial states (7,8). 
As seen from Fig.~\ref{CouIni01}, the initial states in each group will lead to the identical steady states, that is, 
$I_{st,1}=I_{st,2}$, $I_{st,3}=I_{st,4}=I_{st,5}=I_{st,6}$, and $I_{st,7}=I_{st,8}$, where the index $i$ in $I_{st,i}$ denotes the $i$th initial
state of $\hrr$ in Eq.~(55). 
\begin{figure}[!h]
\begin{center}
\includegraphics[scale=0.57]{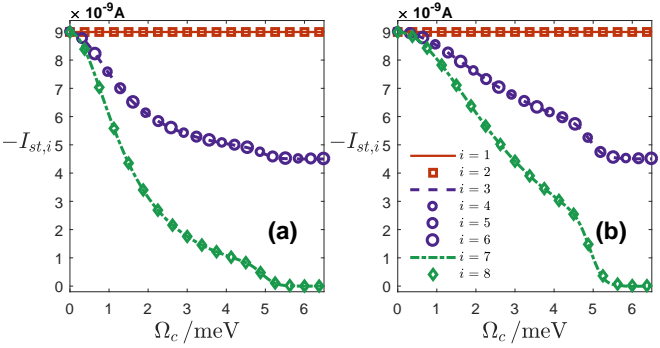}
\end{center}
\caption{The currents $I_{st}$ evaluated from the approximate zeroth order ME, and plotted as a 
function of the energy of the Coulomb interaction $\Om_c$. The commonly shared parameters are set as $\vDe\om =2\pi$Grad/s,
$T=2$K, $\mu_1=35$ meV, $\mu_2=0$, $\Om_1 =\Om_2 =30$ meV, and $\Lam =0.014$. The bandwidths are respectively
set as (a) $b=0.05$; (b) $b=0.1$. The initial conditions of $\hrr$ are set by following Eq.~(\ref{CouIni}).}  \label{CouIni01}
\end{figure}
From Fig.~\ref{CouIni01},
we can see that when $\Om_c=0$ all the initial states lead to the same steady state current.
When $\Om_c\neq0$, there is no Coulomb blockade effect in Group I. However, it is interesting to notice that in Group III, $I_{st}$ is blockaded 
monotonically with respect to $\Om_c$, and it can be blockaded to $0$ when $\Om_c$ is high enough. In Group II, 
$I_{st}$ is also blockaded monotonically, however it can only be blockaded to near half of the original value $I_{st}(\Om_c=0)$.

\begin{figure}[hbt]
\begin{center}
\includegraphics[scale=0.62]{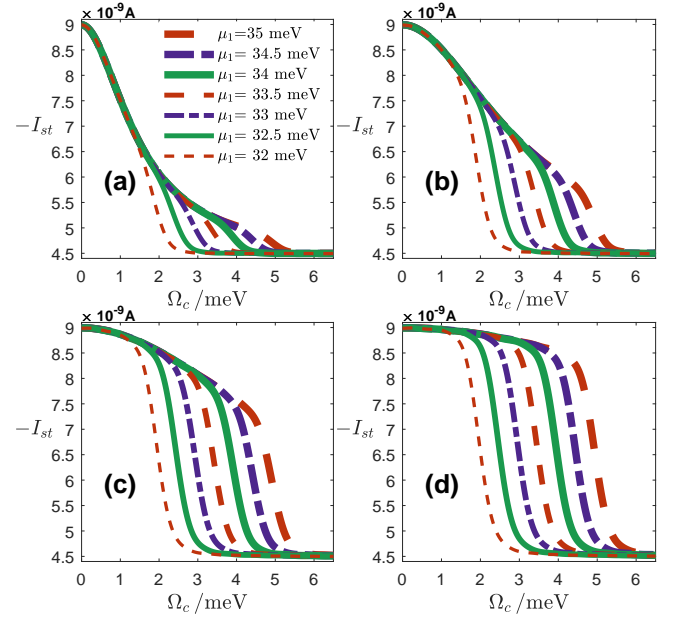}
\end{center}
\caption{The currents $I_{st}$ corresponding to Group II of $\hrr(0)$ evaluated from the approximate zeroth order master
equation, and plotted as a 
function of the energy of the Coulomb interaction $\Om_c$. The parameters are set as $\vDe\om =2\pi$Grad/s,
$T=2$K, $\mu_2=0$, $\Om_1 =\Om_2 =30$ meV, and $\Lam =0.014$. The bandwidths are respectively
set as (a) $b=0.05$; (b) $b=0.1$; (c) $b=0.2$; (d) $b=0.4$.}  \label{CouCorre02}
\end{figure}
\begin{figure}[hbt]
\begin{center}
\includegraphics[scale=0.62]{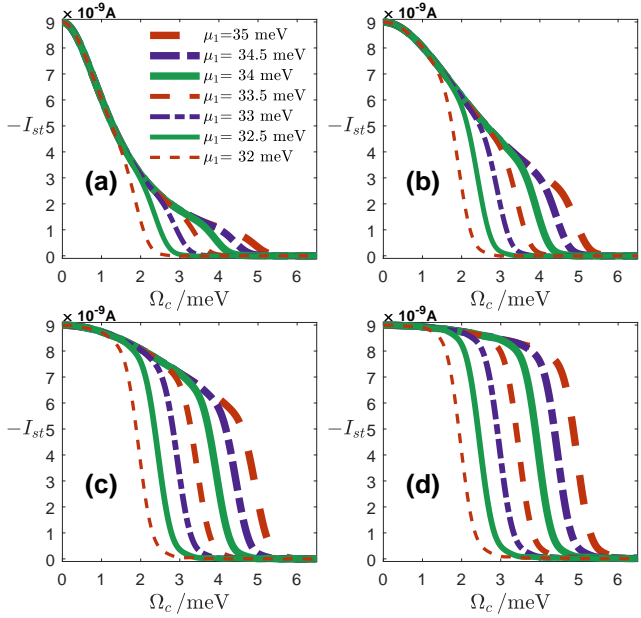}
\end{center}
\caption{The currents $I_{st}$ corresponding to Group III of $\hrr(0)$ evaluated from the approximate zeroth order master
equation, and plotted as a 
function of the energy of the Coulomb interaction $\Om_c$. The parameters are set as $\vDe\om =2\pi$Grad/s,
$T=2$K, $\mu_2=0$, $\Om_1 =\Om_2 =30$ meV, and $\Lam =0.014$. The bandwidths are respectively
set as (a) $b=0.05$; (b) $b=0.1$; (c) $b=0.2$; (d) $b=0.4$.}  \label{CouCorre03}
\end{figure}

In order to get a more detailed picture about the Coulomb blockage for different coulomb coupling energies, 
Figs.~\ref{CouCorre02} and \ref{CouCorre03} are plotted to show how $\Om_c$ influences $I_{st}$ with different chemical potential $\mu_1$ and
the bandwidth $b$. Different from the spin degeneracy case (see Fig.~\ref{Coulomb1current2}),
the blockade is always monotonic with respect to $\Om_c$. 
Particularly, we notice the similarity between the curves in Fig.~\ref{CouCorre02} and curves in Fig.~\ref{CouCorre03}.
In fact, we can show that the currents $I_{st,3}$ and $I_{st,1}+I_{st,7}$ in Fig.~\ref{CouCorre04} are approximately equal,
that is,  $2I_{st,3}=I_{st,1}+I_{st,7}$.  
\begin{figure}[hbt]
\begin{center}
\includegraphics[scale=0.63]{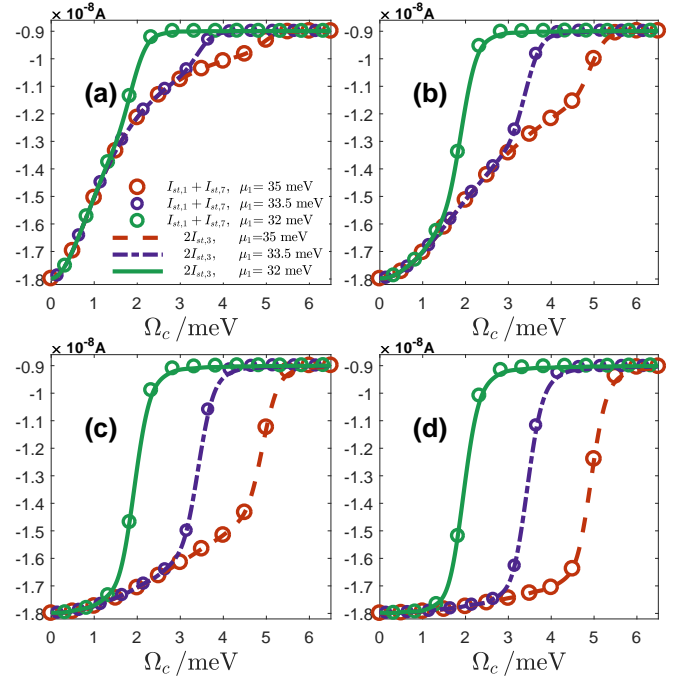}
\end{center}
\caption{The combination relation of the steady state currents which are evaluated from the approximate zeroth order master
equation. The parameters are set as $\vDe\om =2\pi$Grad/s,
$T=2$K, $\mu_2=0$, $\Om_1 =\Om_2 =30$ meV, and $\Lam =0.014$. The bandwidths are respectively
set as (a) $b=0.05$; (b) $b=0.1$; (c) $b=0.2$; (d) $b=0.4$.}  \label{CouCorre04}
\end{figure}

More theoretical analysis of  three groups of initial states can shed light on the steady state classifications.
From Eq.~(\ref{Coulomb2current2}) we see that due to the symmetry of the resonance the currents $I_{1d}$ and $I_{d2}$ 
only involve some elements of $\hrr$: $\rho_{00}$, $(\rho_{11} +\rho_{22})/2$, $\rho_{33}$, and $\rho_{12}^R$. 
According to Eq.~(\ref{Coulomb2ME1}) and (\ref{M}), a set of closed equations can be found for these elements:
\begin {eqnarray}
 \3\6\pa_t
 \begin {bmatrix}
  \rho_{00} \\
  \rho_+ \\
  \rho_{33} \\
  \rho_{12}^R 
 \end {bmatrix}
 =\begin {bmatrix}
  2A   & 2B    & 0    & 2B         \\
  -A  & C\!-\!B    & F   & -C\!-\!B    \\
  0    & -2C    & -2F  & 2C        \\
  -A  & -C\!-\!B    & -F    &  C\!-\!B      
 \end {bmatrix}\e
 \begin {bmatrix}
  \rho_{00} \\
  \rho_+ \\
  \rho_{33} \\
  \rho_{12}^R 
 \end {bmatrix}\e, \q \label{Coulomb2ME2}
\end {eqnarray}
where $\rho_+ \eq (\rho_{11} +\rho_{22})/2$.
Obviously, from Eq.~(\ref{Coulomb2current2}) and (\ref{Coulomb2ME2}), we can derive:
$I_{1d,3}(t)=I_{1d,4}(t)=I_{1d,5}(t)=I_{1d,6}(t)$, and $I_{d2,3}(t)=I_{d2,4}(t)=I_{d2,5}(t)=I_{d2,6}(t)$, 
where $I_{1d,i}$ and $I_{d2,i}$ $(i=1, ..., 8)$ denote the currents corresponding to the eight initial states mentioned in Eq.~(\ref{CouIni}).
Eq.~(\ref{Coulomb2ME2}) can be separated into two parts:
\begin {eqnarray}
 \6\pa_t
 \begin {bmatrix}
  \rho_+ \!-\!\rho_{12}^R \\
  \rho_{33} 
 \end {bmatrix}
 &=&2
 \begin {bmatrix}
  C  & F  \\
  -C & -F 
 \end {bmatrix}
 \begin {bmatrix}
  \rho_+ \!-\!\rho_{12}^R \\
  \rho_{33} 
 \end {bmatrix},  \label{Coulomb2ME3} \\
 \6\pa_t
 \begin {bmatrix}
  \rho_{00} \\
  \rho_+ \!+\!\rho_{12}^R \\
 \end {bmatrix}
 &=&2
 \begin {bmatrix}
  A  & B  \\
  -A & -B 
 \end {bmatrix}
 \begin {bmatrix}
  \rho_{00}  \\
  \rho_+ \!+\!\rho_{12}^R 
 \end {bmatrix}.  \label{Coulomb2ME4}
\end {eqnarray}
When $i=7,8$, the initial states satisfy $\rho_{00}(0)=\rho_+(0) +\rho_{12}^R(0)=0$, then according to the linearity of
Eq.~(\ref{Coulomb2ME4}), we have
$\rho_{00}(t)=\rho_+(t) +\rho_{12}^R(t)=0$. Thus, Eq.~(\ref{Coulomb2ME3}) turns into
\begin {eqnarray}
 \pa_t
 \begin {bmatrix}
  2\rho_+  \\
  \rho_{33} 
 \end {bmatrix}
 =2
 \begin {bmatrix}
  C  & F  \\
  -C & -F 
 \end {bmatrix}
 \begin {bmatrix}
  2\rho_+  \\
  \rho_{33} 
 \end {bmatrix},  \label{Coulomb2ME5}
\end {eqnarray}
and the trace formula of the $\hrr(t)$ turns into $2\rho_+ +\rho_{33}=1$. By combining Eq.~(\ref{Coulomb2ME5}) and 
the trace formula, we can conclude that the four elements of $\hrr(t)$ in Eq.~(\ref{Coulomb2ME2}) corresponding to 
the 7-th and 8-th initial states will converge to the same steady states: 
$\rho_+(t\!\raw\!\ift) =-\rho_{12}^R(t\!\raw\!\ift)=F_{st}/(2(F_{st}-C_{st}))$, $\rho_{33}(t\!\raw\!\ift) =C_{st}/(C_{st}-F_{st})$,
and $\rho_{00}(t)=0$, where the subscript $st$ denotes the steady value of a coefficient. Thus, we have $I_{st,7} =I_{st,8}$.

A similar analysis can be done for Group I with $i=1,2$.  Then $\rho_{33}(t)=0$, and $\rho_+(t)=\rho_{12}^R(t)$ will be obtained.
The steady states in these cases are given by: $\rho_{00}(t\raw\ift) =B_{st}/(B_{st}-A_{st})$,
$\rho_+(t\raw\ift) =\rho_{12}^R(t\raw\ift) =A_{st}/(2(A_{st}-B_{st}))$. So we have $I_{st,1}=I_{st,2}$, and $\rho_{33}(t)=0$.
Since $\rho_{33}$ remains zero for the entire evolution, hence the Coulomb blockade effect does not take place in Group I.

In Eq.~(\ref{CouIni}), we can see that $\hat{\rho}_{r,3}(0)+\hat{\rho}_{r,4}(0)=\hat{\rho}_{r,2}(0)+\hat{\rho}_{r,8}(0)$.
Due to the linearity of the ME and the expressions in Eq.~(\ref{Coulomb2current2}), we can conclude that
$I_{st,3}+I_{st,4}=I_{st,2}+I_{st,8}$. Then, by combining the conclusions about the three groups of the initial states,
the relation $2I_{st,3} =I_{st,1}+I_{st,7}$ can be obtained.

\section{conclusions}
\label{conclusion}

We have applied the NMSSE approach
to derive the ME for the density operator of a quantum dot
coupled to its fermionic environments. We have also derived an approximate ME
 valid to the zeroth order perturbation. We emphasize that the zeroth order perturbation
 effectively ignores the noise terms, but still retains higher order coupling strength terms,
 hence it render the perturbation method a very useful tool in describing a weakly non-Markovian dynamics.
 A higher order noise perturbation beyond the zeroth order approximation is possible if a highly
 non-Markovian environment is involved and more computational resources are available.

With both the exact and approximate methods, the current between the quantum dots and the source and drain 
has been discussed.  The effect of the Coulomb
interaction between the electrons has been discussed on the transport
between the quantum dots and the source and drain. We have found that in the spin degeneracy case 
when $\mu_1$ is sufficiently large, the steady state current may not be monotonically blockaded with respect to the Coulomb energy.
In a specific correlated noise case, we have found that the steady state current may sensitively depend on 
the initial state of the quantum dots.  More specifically, we have shown that there is a class of initial states (Group I states)
for those states the Coulomb blockade does not occur, and for Group III states the maximal Coulomb blockade is observed. All the other
initial states can be described by the combination of Group I and Group II states.

\section*{acknowledgements}
We acknowledge the grant support  from the NSF PHY-0925174,  the National Natural Science
Foundation of China (Grant No. 11304024). We thanks Mr. Quanzhen Ding 
for useful discussions on the manuscript.

\appendix

\section{The algorithm for the evaluation of the coefficients $F_{ij}$ of
the exact ME}\label{ap1}
The coefficients $F_{ij}$ are defined as
\begin {eqnarray}
 F_{ij}(t) \eq \int_0^t ds [K_{b_j}(t,s) u_i^b(t,s) -K_{c_j}^*(t,s) u_i^c(t,s)], \q\q
\end {eqnarray}
where $u_i^\lam$ are the coefficients in the expansions of $\mm(\hq_\lam\hp)$:
\begin {eqnarray}
 \mm(\hq_b(t,s,\bx^*)\hp) \!&=&\! u_1^b(t,s)\hd\hrr(t) +u_2^b(t,s)\hrr(t)\hd,  \non \\
 \mm(\hq_c(t,s,\bx^*)\hp)\+ \!&=&\! -u_1^c(t,s)\hd\hrr(t) -u_2^c(t,s)\hrr(t)\hd.  \q \q
\end {eqnarray}

When the correlation functions satisfy $K_{\lam_j}(t+\ta,s+\ta) =K_{\lam_j}(t,s)$, the coefficients $F_{ij}$ can be constructed
via the basic solution $u$. Here we only list the main results without showing too many details.
We define the correlation functions: $K_{\lam_j}(t-s) \eq K_{\lam_j}(t,s)$, $K_{\lam}(t) \eq \sum_{j=1,2} K_{\lam_j}(t)$,
$K(t) \eq K_b(t) +K^*_c(t)$. $u$ is the solution of the integro-differential equation:
\begin {eqnarray}
 \pa_t u(t) =i\Om u(t) -\int_0^t ds K(s\!-\!t) u(s), \q u(t\!=\!0)\! =\!1. \q\q
\end {eqnarray}

Then $F_{ij}$ can be constructed through the algorithm:
\begin {eqnarray}
  F_{1j} \!&=&\! A_j(t) \!-\!B_j(t) +\fr{C_{1j}(t) \big(1\!-\!D_2(t) \big) \!+\!C_{2j}(t) D_1(t)}{u^*(t)},  \non \\
  F_{2j} \!&=&\! A_j(t) \!-\!B_j(t) -\fr{C_{1j}(t) D_2(t) \!+\!C_{2j}(t) \big(1 \!-\!D_1(t)\big)}{u^*(t)}.  \q \non \\
      \label{4Apd2F}
\end {eqnarray}
$C_{ij}$ can be generated from
\begin {eqnarray}
 C_{1j}(t) &\eq& \int_0^t ds K_{b_j}(t-s) u^*(s), \non \\
 C_{2j}(t) &\eq& \int_0^t ds K^*_{c_j}(t-s) u^*(s).
\end {eqnarray}
$A_j$ and $B_j$ can be obtained from solving equations:
\begin {eqnarray}
  \pa_t A_j(t) &=& C_{1j}(t) C_2^*(t) +u(t)\int_0^t ds K^*_c(t-s) C_{1j}(s),  \non \\
  \pa_t B_j(t) &=& C_{2j}(t) C_1^*(t) +u(t)\int_0^t ds K_b(t-s) C_{2j}(s).  \q\q
\end {eqnarray}
$D_1$ and $D_2$ can be obtained from solving equations:
\begin {eqnarray}
 \6 \pa_t D_1(t) = 2\big( u(t) \sum_j C_{1j}(t) \big)^R, \non \\
  \pa_t D_2(t) = 2\big( u(t) \sum_j C_{2j}(t) \big)^R.  
\end {eqnarray}

When the model contains lots of symmetries, the current can be obtained even without solving the $\hrr$. According to
Eq.~(\ref{Current1D2E}), the average current can be obtained from
\begin {eqnarray}
 I(t)\eq (I_{1d}\!+\!I_{d2})/2 =q_e\big((F_{22}^R \!-\!F_{21}^R)\rho_{00} +(F_{12}^R \!-\!F_{11}^R)\rho_{11}\big).  \non \\ \label{AppI1}
\end {eqnarray}
By utilizing the trace formula of the reduced density operator: $\rho_{00} +\rho_{11}=1$, Eq.~(\ref{AppI1}) can be rewritten as
\begin {eqnarray}
 I/q_e &=& \fr{(F_{22} -F_{21} +F_{12} -F_{11})^R}{2}  \non \\
 && +\fr{(F_{12} -F_{11} -F_{22} +F_{21})^R (\rho_{11} -\rho_{00})}{2}. \q\q\q  \label{AppI2}
\end {eqnarray}
By substituting Eq.~(\ref{4Apd2F}) into the above equation, we get
\begin {eqnarray}
 \!\!\!&&\! F_{12} \!-\!F_{11} \!-\!F_{22} \!+\!F_{21}  \non \\
 \!\!\!&=&\! \fr{C_{12} -C_{11} +C_{22} -C_{21}}{u^*}  \non \\
 \!\!\!&=&\! \fr{\int_0^t ds \big(K_{\!b_2}\!(t\!-\!s) \!+\!K_{\!c_2}^*\!(t\!-\!s) \!-\!K_{\!b_1}\!(t\!-\!s) \!-\!K_{\!c_1}^*\!(t\!-\!s)\big) u^*(s) }
     {u^*(t)}.  \non \\
\end {eqnarray}
And we know that $K_{b_j}+K^*_{c_j}$ is equal to the correlation function in the vacuum case, explicitly
for each $j$, we have
\begin {eqnarray}
 && K_{b_j}(t-s) +K^*_{c_j}(t-s)  \non \\
 &=& \sum_k |t_{jk}|^2 (1-\bar{n}_{jk}) e^{-i\om_k(t-s)} +\sum_k |t_{jk}|^2 \bar{n}_{jk} e^{-i\om_k(t-s)}  
     \non \\
 &=& \sum_k |t_{jk}|^2 e^{-i\om_k(t-s)}.   
\end {eqnarray}
Thus, when $t_{1k} =t_{2k}$, we have $K_{b_1}+K^*_{c_1} =K_{b_2}+K^*_{c_2}$. In such a case,
$F_{12} -F_{11} -F_{22} +F_{21} =0$, and the expression of the average current in Eq.~(\ref{AppI2}) can be simplified:
\begin {eqnarray}
 I(t) =q_e(F_{22} -F_{21} +F_{12} -F_{11})^R/2.
\end {eqnarray}

\end{document}